\pdfoutput=1
\documentclass[journal,onecolumn]{IEEEtran}
\usepackage{cite}
\usepackage{amsmath,amssymb,amsfonts}
\usepackage{algorithmic}
\usepackage{graphicx}
\usepackage{textcomp}
\usepackage[table]{xcolor}
\usepackage{fancyhdr}
\usepackage[hyphens]{url}

\def\BibTeX{{\rm B\kern-.05em{\sc i\kern-.025em b}\kern-.08em
    T\kern-.1667em\lower.7ex\hbox{E}\kern-.125emX}}

\usepackage[normalem]{ulem}
\usepackage[unicode]{hyperref}
\usepackage{color}
\usepackage{soul}
\usepackage{multirow}
\usepackage{caption}
\usepackage{subcaption}

\newcommand\Proximus{Proximu$\$$}
\newcommand{\proximus}{Proximu\$}
\newcommand{\proximusspace}{\proximus\ }

\newcommand{\todo}[1]{ }

\input{ldiffPreamble}
\usepackage{lmodern}
\usepackage{authblk}

\title{\Proximus: Efficiently Scaling DNN Inference in multi-core CPUs through Near-Cache Compute}
    \author[*,+]{Anant V. Nori}
    \author[$\dag$]{Rahul Bera}
    \author[*]{Shankar Balachandran}
    \author[*]{Joydeep Rakshit}
    \author[*]{Om J. Omer}
    \author[**]{\\ Avishaii Abuhatzera}
    \author[**]{Belliappa Kuttanna}
    \author[*]{Sreenivas Subramoney}
    \vspace{-10pt}
    \affil[*]{Processor Architecture Research Lab, Intel Labs, Bangalore}
    \affil[$\dag$]{ETH Zurich, work done while at Intel}
    \affil[**]{Intel Corporation}
    \affil[+]{Corresponding Author : Anant V. Nori, anant.v.nori@intel.com}

\date{}
\begin{document}
\maketitle
\pagestyle{plain}

\vspace{-30pt}
\begin{abstract}
Deep Neural Network (DNN) inference is emerging as the fundamental bedrock for
a multitude of utilities and services. CPUs continue to scale up their raw
compute capabilities for DNN \textit{inference}~\cite{web:intel_clx} along with
mature high performance libraries~\cite{web:mkldnn} to extract optimal
performance.  While general purpose CPUs offer unique attractive advantages for
DNN \textit{inference} at both datacenter~\cite{doc:fbdatacenter} and
edge~\cite{doc:fbedge}, they have primarily evolved to optimize single thread
performance. For highly parallel, throughput-oriented DNN \textit{inference}, this
results in inefficiencies in both power and performance, impacting both raw
performance scaling and overall performance/watt.  

We present \proximus, where we systematically tackle the root inefficiencies in
power and performance scaling for CPU DNN \textit{inference}. Performance scales
efficiently by distributing \textit{light-weight tensor compute near all caches}
in a multi-level cache hierarchy. This \textit{maximizes the cumulative
utilization} of the existing bandwidth resources in the system and
\textit{minimizes movement of data}. Power is drastically reduced through simple
ISA extensions that \textit{encode the structured, loop-y workload behavior}.
This enables a bulk offload of pre-decoded work, with loop unrolling in the
light-weight near-cache units, effectively \textit{bypassing the power-hungry
stages} of the wide Out-of-Order (OOO) CPU pipeline.

Across a number of DNN models, \proximusspace achieves a 2.3$\times$ increase in
\texttt{convolution} performance/watt with a 2$\times$ to 3.94$\times$ scaling
in raw performance. Similarly, \proximusspace achieves a 1.8$\times$ increase
in \texttt{inner-product} performance/watt with 2.8$\times$ scaling in
performance. With no changes to the programming model, no increase in cache
capacity or bandwidth and minimal additional hardware, \proximusspace enables
unprecedented CPU efficiency gains while achieving similar performance to state-of-the-art
Domain Specific Accelerators (DSA) for DNN \textit{inference} in this AI era.

\end{abstract}

\vspace{-15pt}
\section{Introduction}
\label{sec:intro}

New data-centric paradigms of compute have made machine learning (ML) and Deep
Neural Networks (DNN) pervasive in all fields of human endeavor. The race to
build the optimal hardware for DNN execution continues with custom Domain
Specific Accelerators (DSA), programmable FPGAs and general purpose GPUs and
CPUs all throwing their hats in the ring.
CPUs offer unique attractive advantages for DNN-\textit{inference} in the
datacenter~\cite{doc:fbdatacenter} and also at the edge~\cite{doc:fbedge}.
Advances in DNN-inference topologies and algorithms continue at a rapid pace. The
programmable general-purpose nature of CPUs with their rich and mature
ecosystem of tools and programming models allows for implementation of
functionality that is not present in custom DNN hardware~\cite{smartphone},
enabling quick development and deployment.
Additionally, DNN topologies do not exist in a vacuum and a tight coupling of
DNN and non-DNN tasks is required to meet strict inference latency requirements
for sufficient quality-of-service to the end users. 
Software driver-based offload of DNN-inference tasks to a separate piece of
hardware (DSAs, GPUs) incurs unacceptable latency and memory costs. Hence CPUs
are better suited to real time DNN-inference tasks~\cite{mark,mobile_cv}.
DNN tasks with limited parallelism, like Recurrent NNs, fit more naturally to
CPUs which have few fast cores, than to GPUs which have many slow
cores~\cite{deepcpu}.
Finally, the wide prevalence of CPUs already in datacenters provisioned for peak
load levels in conjunction with diurnal load cycles, leads to the abundant
availability of ``free" CPU compute for DNN-inference~\cite{doc:fbdatacenter}.
Hence, efficient scaling of DNN-inference on CPUs is highly
crucial both for meeting customer Service-Level Agreements(SLAs) and enabling Total Cost of Ownership (TCO) savings for datacenter
providers as applications using DNNs proliferate.

\begin{figure}
    \centering
    \begin{minipage}{.25\textwidth}
    	\centering
    	\includegraphics[width=\linewidth,height=1.8in]{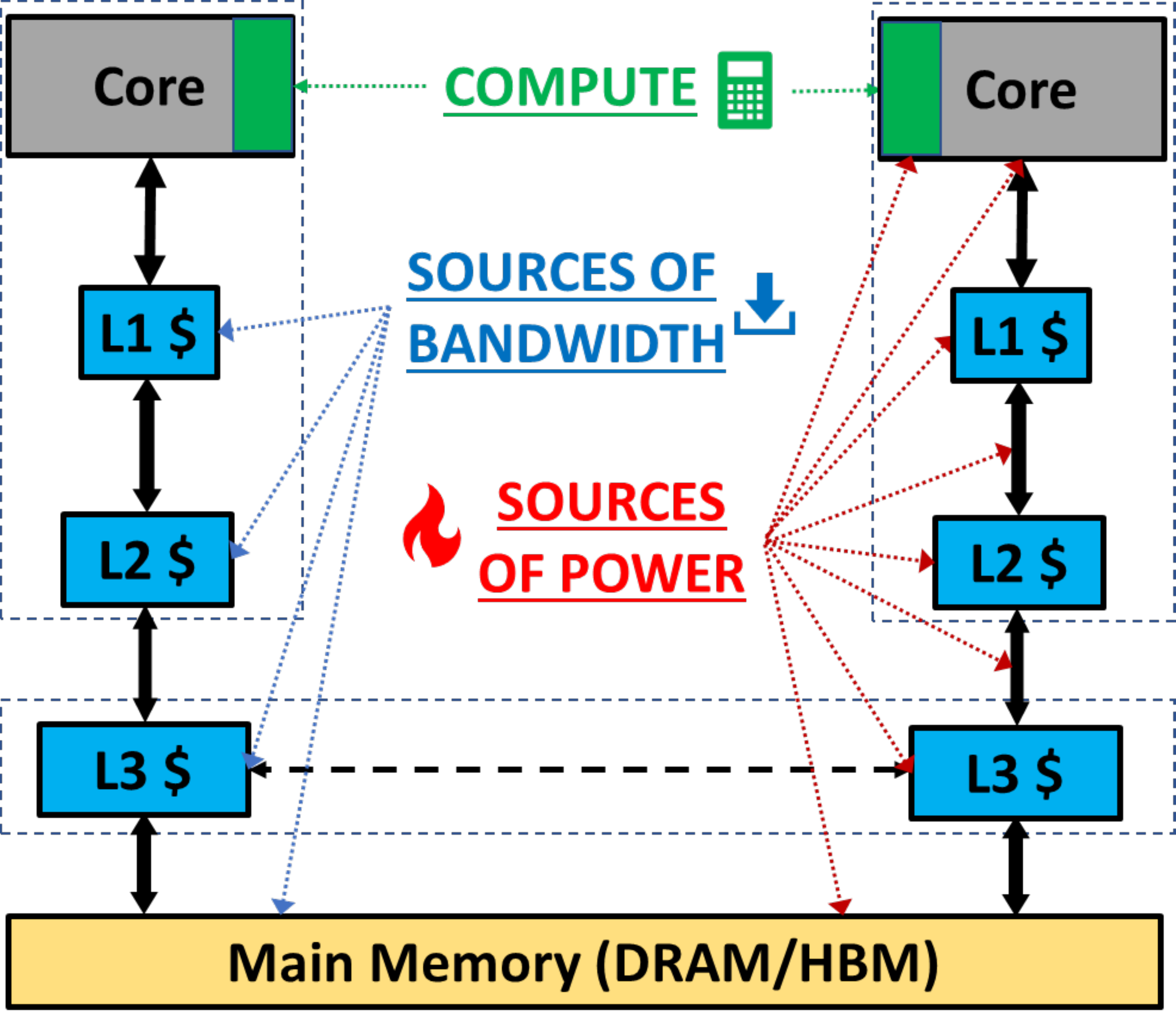}
        \captionof{figure}{Modern CPUs with multi-level memory hierarchies}
    	\label{fig:intro_fig_with_power}
    \end{minipage}
    \begin{minipage}{.25\textwidth}
    \centering
	\includegraphics[width=\linewidth,height=1.8in]{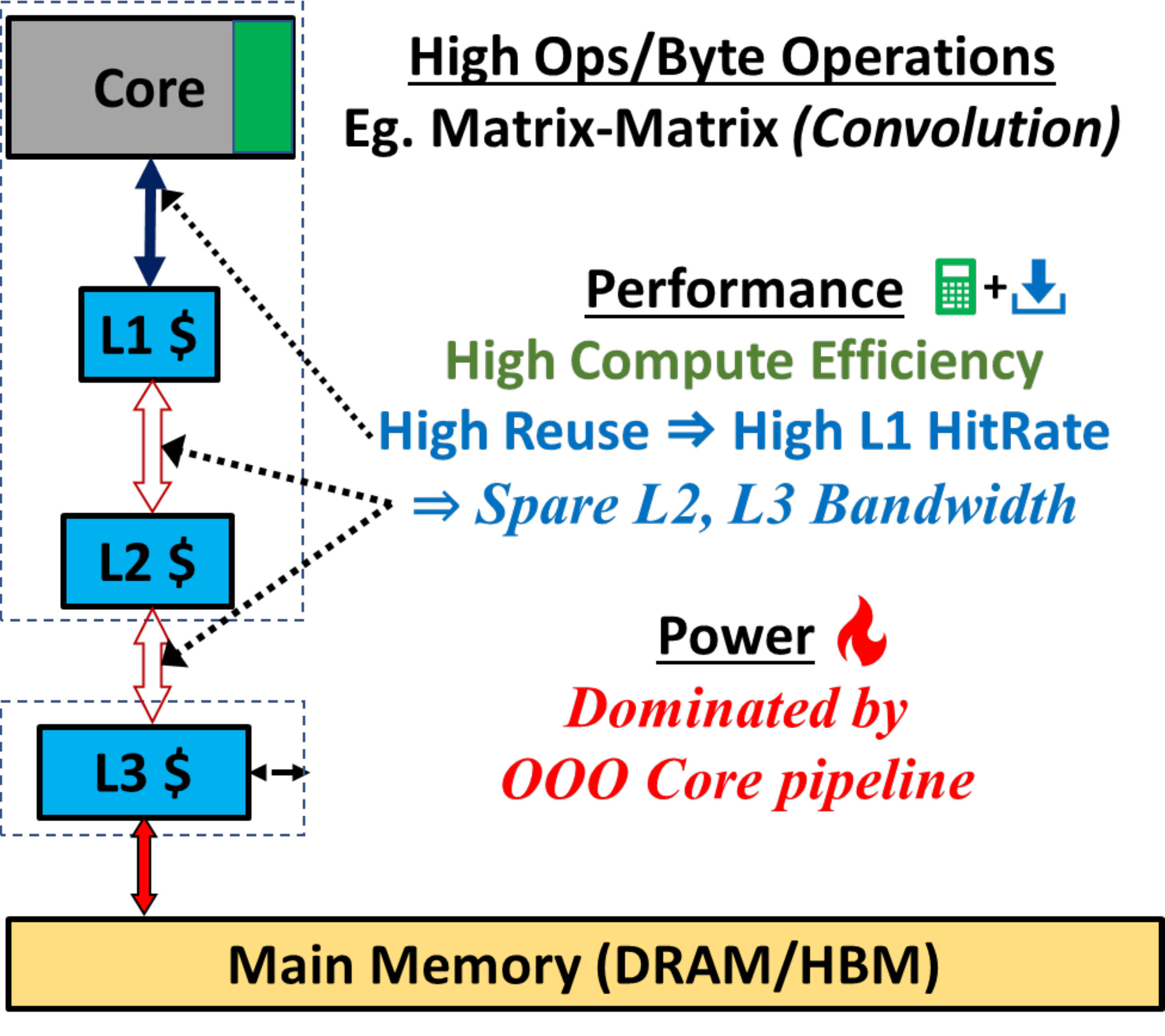}
    \captionof{figure}{High Ops/Byte Matrix-Matrix Primitives}
	\label{fig:conv_intro3}
    \end{minipage}
    \begin{minipage}{.25\textwidth}
	\centering
	\includegraphics[width=\linewidth,height=1.8in]{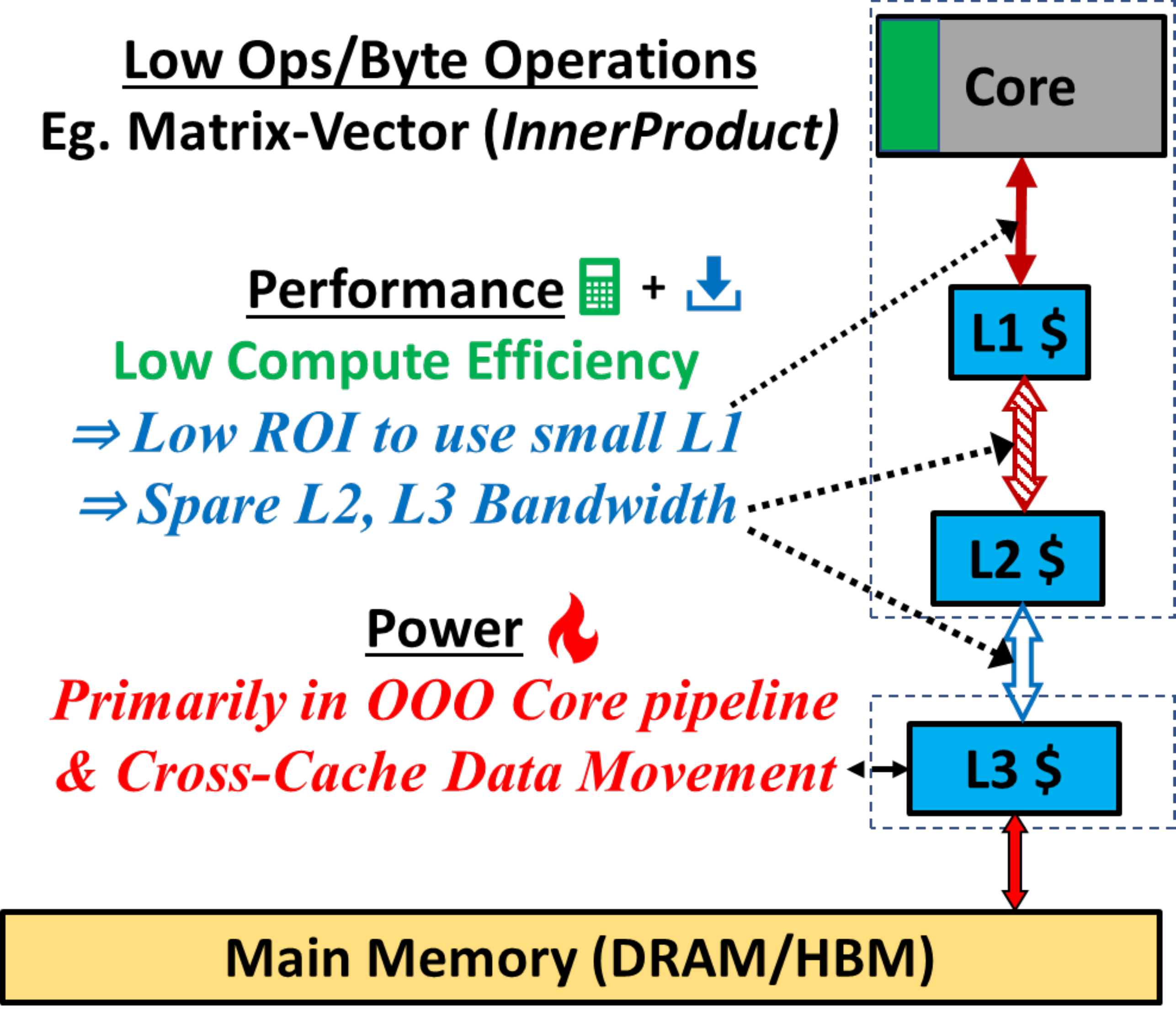}
    \captionof{figure}{Low Ops/Byte Matrix-Vector Primitives}
	\label{fig:fc_intro3}
    \end{minipage}
    \vspace{-20pt}
\end{figure}

However, modern CPUs have evolved to optimize primarily for
single thread performance. The conventional CPU organization
(Figure~\ref{fig:intro_fig_with_power}) involves wide and deep OOO cores, with
all compute placed ``monolithically" atop a serially accessed multi-level
cache hierarchy designed to minimize average load latency. While each cache
level is potentially an independent source of bandwidth, all loads and stores
must go through the L1 cache, thus restricting it to be the primary source of
bandwidth.  Furthermore, every instance of every instruction across all loop
iterations in the workload is unrolled and travels through the entire CPU
pipeline, consuming a significant amount of power. 

DNN-inference primitives like \texttt{convolution, inner-product} are highly
structured and repetitive, with any invocation having a number of fixed
iteration count loops. They are also heavily data-parallel with performance governed
by raw compute throughput and a required data throughput (bandwidth) to feed it.
Generational compute scaling in CPUs is achieved via both intra-core
scaling~\cite{web:avx512,web:intel_clx} and multi-core scaling. The required
bandwidth to feed the compute however, depends on primitives themselves; the
higher the compute intensity (Ops/Byte) the lower the bandwidth required, and
vice-versa. As DNN usages and topologies evolve, there is an increasing
heterogeneity in their Ops/Byte (Compute/Bandwidth) requirements.

The monolithic core-centric CPU organization results in sub-optimal performance,
resource utilization and power when executing DNN primitives with diverse
Ops/Byte requirements. Matrix-Matrix primitives (like convolution) illustrated
in Figure~\ref{fig:conv_intro3} typically have high Ops/Byte. High reuse and
hit-rate in the L1 cache delivers sufficient bandwidth to achieve close to peak
compute efficiency. However, serialized accesses through the hierarchy means
the L2 and L3 cache bandwidths are heavily underutilized.  Matrix-Vector
primitives (like inner-product), shown in Figure~\ref{fig:fc_intro3} have much
lower Ops/Byte. Low hit-rates and bandwidth from small L1 caches
result in low compute efficiencies. In fact, any data movement into the L1 is
essentially wasteful and puts unnecessary pressure on outer cache levels while
consuming power. Larger L2 caches have higher hit-rates resulting in spare L3
bandwidth. \textit{Power consumption is dominated by the unrolled fetch, decode,
allocate and dispatch of every instruction despite the repetitive structured 
nature of execution}. 

We present \proximus, where we systematically tackle the root inefficiencies in
power and performance scaling for CPU DNN inference. Performance scales
efficiently by distributing \textit{light-weight tensor compute near all caches}
in a multi-level cache hierarchy. This \textit{maximizes the cumulative
utilization} of the existing bandwidth resources in the system and
\textit{minimizes unnecessary movement of data across the hierarchy}. We
leverage the structured and repetitive nature of DNN-inference primitives to
define simple ``Proximity Support Extensions" (PSX) to the ISA that condense and
encode multiple levels of loops. The core does fewer fetches and decodes, with a
bulk offload of decoded tensor work (load/store/compute) to the light-weight
near cache ``Tensor Functional Units" (TFU). With unrolled tensor execution
within the TFU, a majority of the power-hungry stages of the OOO CPU pipeline are
effectively bypassed, drastically reducing power. We further leverage the PSX
extensions and Simultaneous Multi Thread (SMT) capabilities of the CPU to
distribute work across threads and cores with no changes to the CPU programming
or memory model.  

We make the following key contributions in this paper.
\begin{itemize}
\item We do a fundamental analysis of state-of-the-art implementations of
multiple DNN-inference primitives executed on state-of-the-art datacenter CPUs,
and identify key bottlenecks to performance scaling and performance/watt efficiency.
\item We present \textbf{\proximus}, where we \textit{distribute} light-weight
Tensor Functional Units, \textit{near each level of cache}. \proximusspace
\textbf{maximizes efficient utilization of the existing cumulative bandwidth} in
the system and \textbf{minimizes data movement}. \proximusspace adds minimal
additional hardware with \textit{no increase in cache capacity or bandwidth} to the CPU,
while \textbf{scaling performance to levels matching state-of-the-art DNN DSAs}.
\item We develop simple ``Proximity Support Extensions" to the ISA that condense
and encode multiple loops of fixed iteration count information. This effectively
ensures that \textbf{unnecessary power-hungry stages of the OOO CPU pipeline are effectively
bypassed} and all work (unrolling and execution) is performed in close proximity to the
data, drastically improving achieved performance/watt.
\item By leveraging the PSX extensions and existing SMT capabilities of cores,
\proximusspace requires \textbf{no change to the CPU programming or memory
model}.
\end{itemize}

Evaluated across multiple DNN models, \textbf{\proximusspace achieves a
2.3$\times$ improvement in \texttt{convolution} performance/watt with a
2$\times$ to 3.94$\times$ scaling in raw performance}. Similarly,
\textbf{\proximusspace achieves a 1.8$\times$ increase in \texttt{inner-product}
performance/watt with 2.8$\times$ performance}. With \textit{no changes to the
programming model, no increase in cache capacity or bandwidth} and minimal
additional hardware, \textbf{\proximusspace enables unprecedented CPU efficiency gains
and TCO savings for datacenters while matching performance levels of
state-of-the-art DNN DSAs}.

\newpage
\section{Characterization and Opportunity}
\label{sec:background}

We first perform an in-depth power and performance characterization of multiple
primitives common to DNN inference on state-of-the-art CPU configurations. The
goal is to derive insights that can lead to efficient performance and
performance/watt scaling. 

\subsection{Programming and Execution Model}
\begin{figure*}[h]
	\centering
	\includegraphics[width=1.0\linewidth]{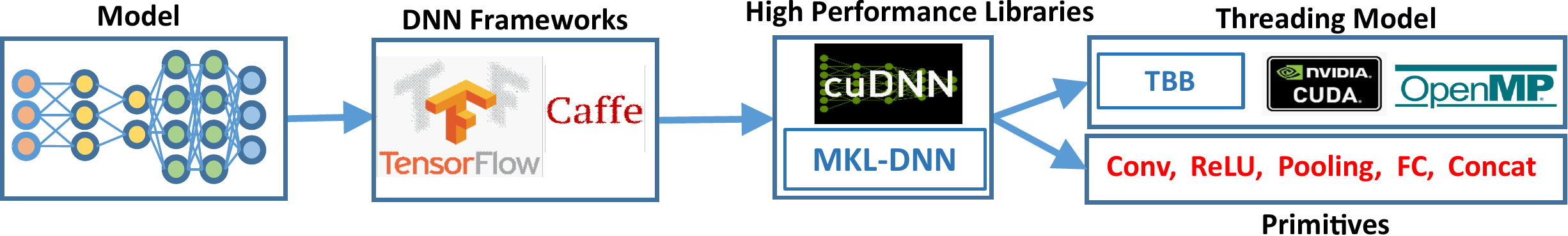}
	\caption{Software stack for deep neural network tasks}
	\label{fig:flow_overview}
\end{figure*}

\begin{figure}[h]
	\centering
	\includegraphics[width=0.6\linewidth]{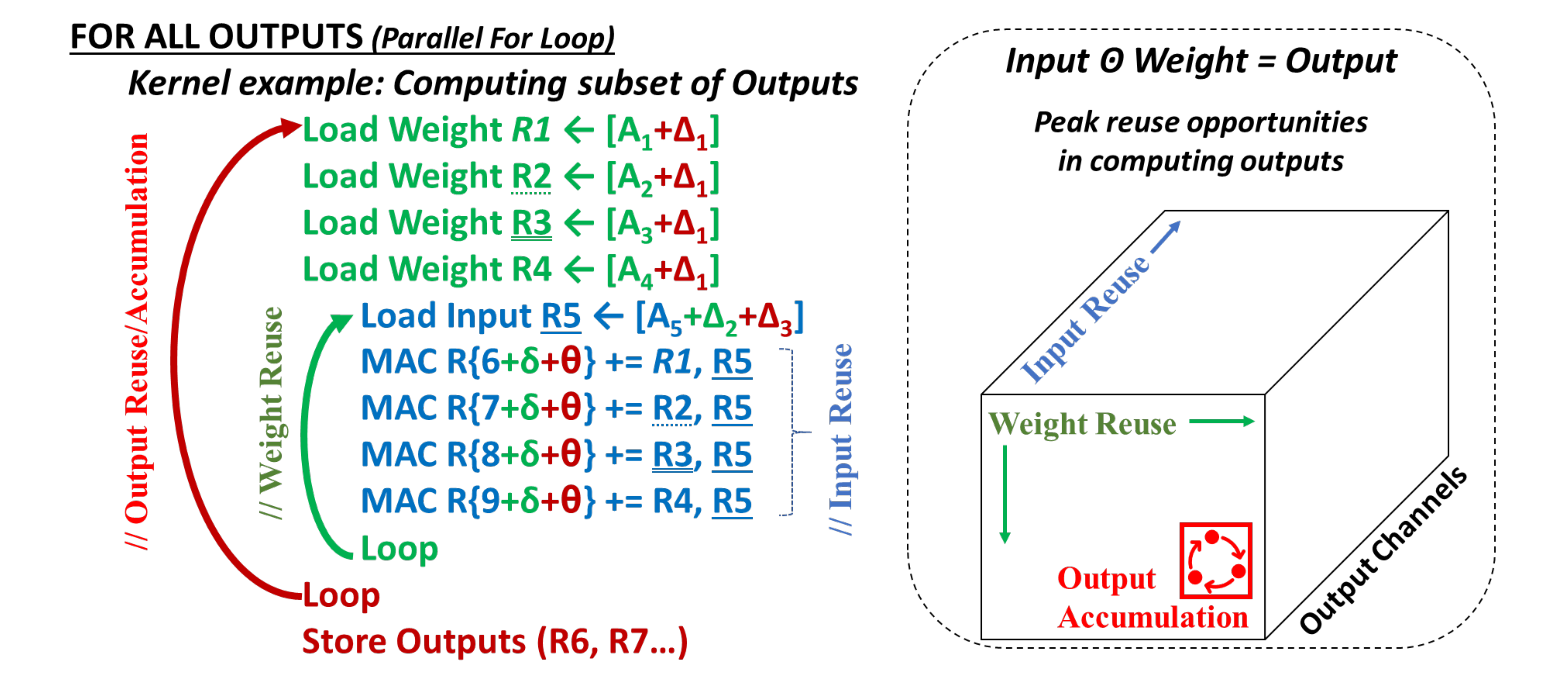}
	\caption{1x1 Convolution example: Kernel loops and data reuse vs. peak
	theoretical reuse}
	\label{fig:repetitive_loops}
\end{figure}

Figure~\ref{fig:flow_overview} depicts the overall flow currently followed when
implementing DNN models. Models are specified in frameworks like
Tensorflow~\cite{web:tensorflow}, Caffe~\cite{web:caffe},
PyTorch~\cite{web:pytorch}, MXNet~\cite{web:mxnet}, OpenVino~\cite{web:openvino}
etc. These frameworks provide developers with easy-to-use APIs to describe
topologies and model parameters while abstracting away the underlying hardware.
The frameworks in-turn leverage highly-optimized, platform-specific libraries
like Intel MKL-DNN~\cite{web:mkldnn}~\cite{web:tfmkldnn}~\cite{web:caffemkldnn},
AMD's BLIS or libFLAME~\cite{web:amd_library}, ARM's compute
libraries~\cite{web:arm_library} and Nvidia cuDNN~\cite{web:cudnn} to extract
maximum performance from the underlying hardware. The inner-most loops of DNN
primitives are typically implemented in a vectorized, highly optimized (often
JITed~\cite{doc:anatomy_paper}) manner for optimal performance and maximum data
reuse (Figure~\ref{fig:repetitive_loops}). The outer-most loops in the
primitives are parallelized using well established threading run-times
(OpenMP~\cite{web:openmp}, TBB~\cite{web:tbb}) to distribute work across compute
cores in the target hardware.

\subsection{Performance and Power Analysis}

We evaluate state-of-the-art MKL-DNN \cite{web:mkldnn} primitives (up to 198X
performance improvement~\cite{web:mkldnn198X}) on
state-of-the-art CPU configurations across multiple DNN
topologies\footnote{Our evaluation methodology and workloads are described in
detail in Section~\ref{sec:eval}.}. \textit{We focus on int8 data types since 
several seminal studies ~\cite{doc:icml_precision} have shown that 8 bit (or
lower) precision is sufficient for inference accuracy.} \textit{The use of lower
precisions reduces dependence on expensive off-chip DRAM bandwidth with the
\textbf{remaining bandwidth bottlenecks becoming mostly on-die thereafter}.}

The compute intensity or Ops/Byte property of any DNN primitive plays a
fundamental role in determining its bandwidth requirements (and hence compute
efficiency) from the system. This metric needs to be evaluated at multiple
levels of abstraction:
\begin{itemize}
\item \textbf{\textit{Algorithm:}} The theoretical peak Ops/Byte is based on the
work being performed if we had an ``infinite" register file (RF) holding data.
For example, in 1x1 \texttt{convolution} (Figure~\ref{fig:repetitive_loops}),
every weight element is reused across all input plane
elements in the same input channel (for different output plane elements).
Similarly, input and output elements also have peak possible reuse
opportunities (shown in Figure~\ref{fig:repetitive_loops}). 
\item \textbf{\textit{Kernel:}} The software implementation that extracts reuse
out of the finite RF of the compute core and depends on the RF size and the data-flow
implemented. This determines the minimum number of loads and
stores to be executed. For the kernel example in
Figure~\ref{fig:repetitive_loops}, the number of weight loads (reused across
inputs in the inner-most loop) and the number of iterations of the innermost
loop (reusing weights to compute different outputs) is governed by the RF size.
\item \textbf{\textit{Hardware:}} The kernel execution on the underlying
hardware, where on-die cache hit-rates determine bandwidth delivery to the core
and the cross-cache data movement incurred. We define \texttt{Data Movement Overheads}
introduced by the hardware as the ratio of cumulative cross-cache data movement
(fills and evictions) to the loads and stores to/from the compute core's RF (determined by the
kernel).
\end{itemize}

\begin{table}[t!]
  \centering
  \caption{ResNet-50 Convolution and Transformer Inner Product Characterization}
    \begin{tabular}{|p{3.8em}|c|c|c|c|c|c|}
    \hline
    \multicolumn{1}{|c|}{} & \multicolumn{3}{p{8.145em}|}{\textbf{ResNet50 \newline{}Convolution}} & \multicolumn{3}{p{7.645em}|}{\textbf{Transformer \newline{}InnerProduct}} \\
    \hline
    \multicolumn{1}{|c|}{\textbf{Metric}} & \textbf{Avg.} & \textbf{Min} & \textbf{Max} & \textbf{Avg.} & \textbf{Min} & \textbf{Max} \\
    \hline
    \rowcolor[rgb]{ 0,  0,  0} \multicolumn{7}{|c|}{\textcolor[rgb]{ 1,  1,  1}{\textbf{Ops/Byte: Based on Algorithm}}} \\
    \hline
    \multicolumn{1}{|c|}{\textbf{Input}} & 1021  & 32    & 4608  & 1727  & 1024  & 33708 \\
    \hline
    \multicolumn{1}{|c|}{\textbf{Weight}} & 2245  & 100   & 25600 & 1     & 1     & 1 \\
    \hline
    \multicolumn{1}{|c|}{\textbf{Output}} & 998   & 64    & 4608  & 2057  & 1024  & 33708 \\
    \hline
    \rowcolor[rgb]{ 0,  0,  0} \multicolumn{7}{|c|}{\textcolor[rgb]{ 1,  1,  1}{\textbf{Memory Transactions/Op-Instr: Based on Kernel}}} \\
    \hline
    \multicolumn{1}{|c|}{\textbf{Loads}} & 0.49  & 0.39  & 0.59  & 1.35  & 1.19  & 1.41 \\
    \hline
    \multicolumn{1}{|c|}{\textbf{Stores}} & 0.058 & 0.003 & 0.25  & 7E-04 & 3E-05 & 9E-04 \\
    \hline
    \rowcolor[rgb]{ 0,  0,  0} \multicolumn{7}{|c|}{\textcolor[rgb]{ 1,  1, 1}{\textbf{Hardware: Performance (Peak 2*64 MAC/Cycle)}}} \\
    \hline
    \multicolumn{1}{|c|}{\textbf{Ops/Cyc}} & 120.4  & 100.0  & 127  & 12.99 & 8.81  & 14.02 \\
    \hline
    \rowcolor[rgb]{ 0,  0,  0} \multicolumn{7}{|c|}{\textcolor[rgb]{ 1,  1,  1}{\textbf{Hardware: On-Die Cache Hit-Rate}}} \\
    \hline
    \multicolumn{1}{|c|}{\textbf{L1 \$}} & 86\%  & 57\%  & 98\%  & 23\%  & 15\%  & 26\% \\
    \hline
    \multicolumn{1}{|c|}{\textbf{L2 \$}} & 88\%  & 51\%  & 99\%  & 72\%  & 0\%   & 100\% \\
    \hline
    \multicolumn{1}{|c|}{\textbf{L3 \$}} & 99.4\% & 97.6\% & 99.9\% & 99\%  & 64\%  & 100\% \\
    \hline
    \rowcolor[rgb]{ 0,  0,  0} \multicolumn{7}{|c|}{\textcolor[rgb]{ 1,  1,  1}{\textbf{Hardware: Data Movement Overhead}}} \\
    \hline
    \multicolumn{1}{|c|}{\textbf{L1-L2}} & 20\%  & 1\%   & 69\%  & 109\% & 81\%  & 121\% \\
    \hline
    \multicolumn{1}{|c|}{\textbf{L2-L3}} & 2\%   & 0.3\% & 5.8\% & 47\%  & 27\%  & 99\% \\
    \hline
    \multicolumn{1}{|c|}{\textbf{Total}} & 22\%  & 2\%   & 71\%  & 156\% & 147\% & 181\% \\
    \hline
    \end{tabular}%
  \label{table:tab_ci}%
\end{table}%

\subsubsection{\texttt{Convolution} Characterization}
\label{sec:csize}

Convolution is essentially a high Ops/Byte matrix-matrix operation. We evaluate
the optimized MKL-DNN option that allows for fusing of the convolution and ReLU
(non-linear function on the output) primitives. Table~\ref{table:tab_ci} details
our characterization of the convolution primitive across all convolutional
layers of ResNet-50~\cite{doc:resnet50} and shows several interesting insights.
First, the MKL-DNN convolution kernels \textit{subsume most of the variability
in Ops/Byte across layers} through reuse within the core's RF.
The kernels employ output-stationary data-flows, requiring very low store
bandwidth (Stores/MAC-Instr). \textit{Interestingly}, input and weight reuse
variability is also subsumed within the RF resulting in a fairly steady 0.5
Loads/MAC-Instr requirement across all layers. Second, high average L1 hit-rates (86\%)
result in high compute efficiency (120 MACs/cycle out of a peak 128 (Intel
CascadeLake cores have two 64 MAC/cycle execution units per core)).  Fills and
evictions at the L1 cache still add an average 20\% overhead in data movement.
The \texttt{conv1} layer has poor L1 hit-rate, adding 69\% data movement
overhead at L1, dropping performance to 50 MACs/cycle. 

\textit{Performance Opportunity:} High L1 hit-rates coupled with the 0.5
loads/MAC-instr requirements means that we use only about 60\% of available L1
bandwidth (2 loads/cycle/core at L1 for Intel Cascadelake). \textit{Furthermore,
the L2 and L3 bandwidths are still hugely under-utilized}. This can be
exploited by placing tensor compute near each of these caches, enabling further
scaling of performance without any increase in overall on-die capacity or
bandwidth. More re-use directly from these caches would also reduce data
movement to the L1 caches. The 0.5 loads/MAC-instr requirement and peak
bandwidths of each cache determines the peak compute required near each cache level.

\subsubsection{\texttt{Inner-Product} Characterization}

Inner-Product primitives involve matrix-vector operations and have lower peak
Ops/Byte compared to convolutions. These primitives are prominent in Recurrent
DNN models and Sequencer-to-Sequence models (eg. Transformer~\cite{transformer}) which are heavily used in applications like
natural language processing. Table~\ref{table:tab_ci} also details our
inner-product characterization for all layers in Transformer. This primitive has
a poor 23\% L1 hit-rate, which coupled with a high 1.35 Load/MAC-instr bandwidth
requirement results in low compute efficiency achieving only 13 MACs/cycle.
Furthermore, we see up to a whopping 156\% overhead in cross-cache data
movement.

\textit{Performance Opportunity:} While L1 hit-rates are low, hit-rates in the
larger L2 (1MB) and L3 (1.375MB per core) are significantly better. {\em Tensor
compute placed directly near these caches, bypassing the small L1 entirely,
would leverage the higher hit-rates for higher bandwidth to feed
the compute}. Along with eliminating all data movement to an under-sized L1, we
would see better performance, with no increase in cache capacity or
bandwidth.

\subsubsection{Pooling/Concat}

We also evaluate Pooling (dimensionality reduction) and Concat primitives and
they mainly involve data movement with low data reuse. For example, models like
DenseNet-169~\cite{doc:densenet} pass lower level features (outputs) directly to
later layers as inputs, using the Concat primitive to prepare data.
Near L2 and/or L3 caches execution would reduce this data movement cost.

\subsubsection{Power Analysis}

\begin{figure}[h]
	\centering
	\includegraphics[width=0.5\linewidth]{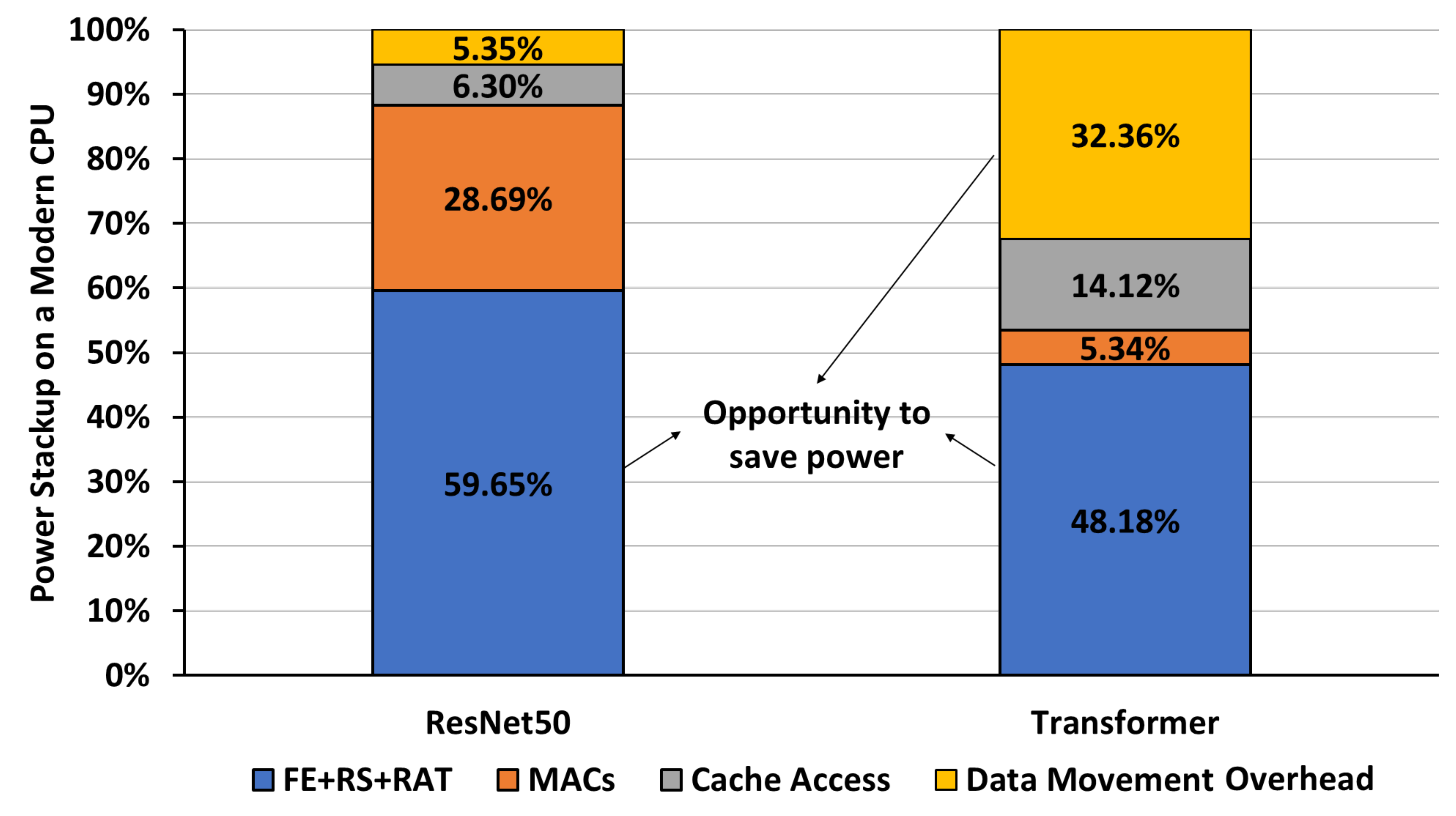}
	\caption{Stackup of Power Consumption in Convolution Dominated
    ResNet50 and Inner Product Dominated Transformer}
	\label{fig:background_energy_stackup}
\end{figure}

Figure~\ref{fig:background_energy_stackup} shows the contribution to total power
from various clusters in the CPU. CPUs unroll \textit{\underline{every
instance}} of \textit{\underline{every instruction}} in \textit{\underline{each
iteration}} of \textit{\underline{all loops}} into \textit{\underline{every
stage}} of the CPU pipeline.  Despite the structured, fixed iteration count
loopy nature of the DNN kernels (Figure~\ref{fig:repetitive_loops}) all
instructions go through fetch and decode, allocation and dispatch. \textbf{Register
allocation and renaming (RAT) and OOO dispatch (from RS) is extremely expensive
in power for wide and deep OOO cores}. For compute bound ResNet-50, these stages contribute to 60\%
of total power! For bandwidth bound inner-product primitives in Transformer,
they contribute to 50\% of total power with cache and data movement adding
another 45\%. 

\textit{Power Opportunity:} With structured and repetitive DNN kernels, loop
unrolling should happen in a ``lean" scheduler, close to the tensor compute.
Using a ``macro"-ISA that encodes this loop information, the CPU can offload
multiple loops of decoded (but not unrolled) work to the ``lean", low-cost
near-cache compute \textbf{effectively bypassing power-hungry stages of the
legacy CPU pipeline} for most
of the execution.

\subsubsection{Summary}

Table~\ref{table:tab_summary} concludes this section by summarizing our main
observations for performance and power. A heterogeneity in characteristics across various
primitives leads to sub-optimal performance and/or resource utilization in the
CPU. Optimally executing primitives near caches best suited to their
requirements can provide performance, power and resource utilization benefits.
Furthermore, we should leverage the structured, fixed iteration count loop
nature of the workloads to bypass or minimize usage of the power-hungry
front-end stages of the legacy CPU pipeline - preferably unrolling and scheduling
pre-decoded instructions near the execution units.
\begin{table*}[h]
	\centering
	\caption{Primitive Characterization Summary}
	\footnotesize
    \setlength\tabcolsep{2pt}
	\begin{tabular}{|p{10em}c|cc|}
    \hline
	\multicolumn{1}{|c|}{\textbf{Primitive}} & \textbf{Observations} & \multicolumn{1}{p{7.5em}|}{\textbf{Data Movement Overhead}} & \textbf{Opportunity} \\
    \hline
	\rowcolor[rgb]{ .851,  .882,  .949} \multicolumn{1}{|c|}{\textbf{Convolution}} & \multicolumn{1}{p{20.145em}|}{Bandwdith over-provisioned w.r.t. compute,\newline{}Under-utilization of L2/L3 Bandwidth} & \multicolumn{1}{p{7.5em}|}{Mostly at \newline{}L1-L2} & \multicolumn{1}{p{20em}|}{Perform tensor compute near \textbf{all} caches \newline{}(L1,L2,L3)} \\
    \hline
	\rowcolor[rgb]{ .851,  .882,  .949} \multicolumn{1}{|c|}{\textbf{Inner-product}} & \multicolumn{1}{p{20.145em}|}{Compute over-provisioned w.r.t. bandwidth,\newline{}Poor hitrate at 32KB L1} & \multicolumn{1}{p{7.5em}|}{High at L1-L2 \newline{}and L2-L3} &  \multicolumn{1}{p{20em}|}{Place tensor compute near \textbf{large} caches \newline{}(L2 and L3)} \\
    \hline
	\rowcolor[rgb]{ .851,  .882,  .949} \multicolumn{1}{|c|}{\textbf{Pooling/\newline{}Concat}} & \multicolumn{1}{p{20.145em}|}{Low data reuse\newline{}Mostly data movement} & \multicolumn{1}{p{7.5em}|}{High at L1-L2 \newline{}and L2-L3} &  \multicolumn{1}{p{20em}|}{Execute near \textbf{outer} cache levels \newline{}(L3/L2)} \\
    \hline
	\rowcolor[rgb]{ .973,  .796,  .678} \multicolumn{2}{|p{30.145em}|}{Power dominated by unrolled wide OOO Fetch, Alloc and Dispatch} & \multicolumn{2}{c|}{\textbf{Exploit structured/loopy kernel to encode multiple loops}} \\
    \hline
	\end{tabular}%
	\label{table:tab_summary}%
\end{table*}%

\newpage
\section{Proximu$\$$}
\begin{figure}[h]
  \centering
  \includegraphics[width=0.5\linewidth]{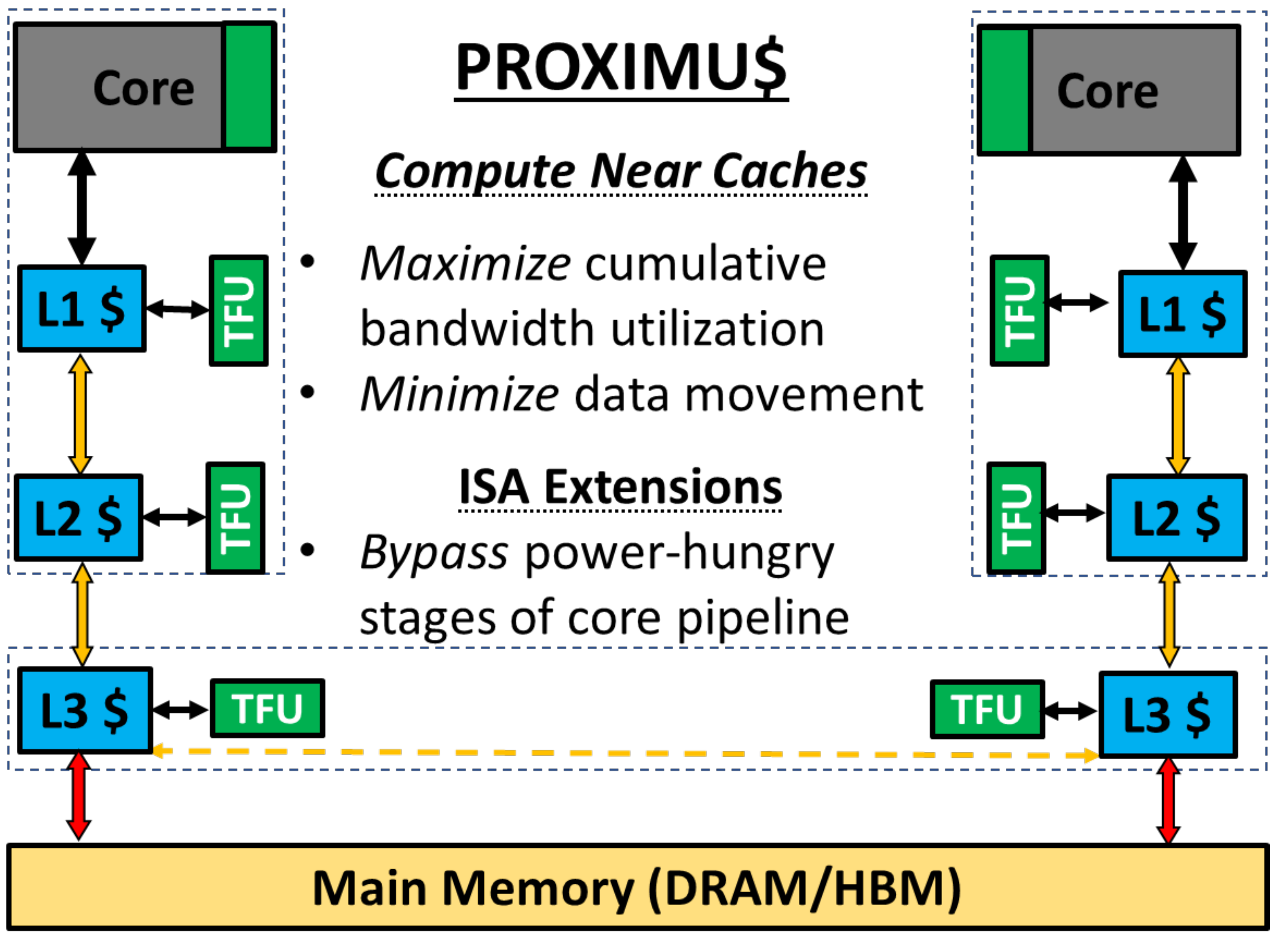}
  \caption{An overview of \proximusspace design}
  \label{fig:design_overview}
\end{figure}

We present \proximus, which \textit{places light-weight ``Tensor Function
Units" (TFU) near \textbf{all} on-die caches in the system} as depicted in
Figure~\ref{fig:design_overview}. In combination with ``Proximity Support
Extensions" (PSX) to the ISA, the goal is to efficiently leverage existing
system resources (maximizing or minimizing use as required) for performance and
power benefits. Crucially, \proximusspace also retains the existing CPU
programming and memory models which can significantly speed up development and
deployment efforts. We now detail the architectural, micro-architectural and
programming model aspects of \proximus.

\subsection{\proximus: Architectural Support}
\label{sec:des_arch}
\subsubsection{Proximity Support Extensions~(PSX)}

\begin{figure}[h]
  \centering
  \includegraphics[width=0.6\linewidth]{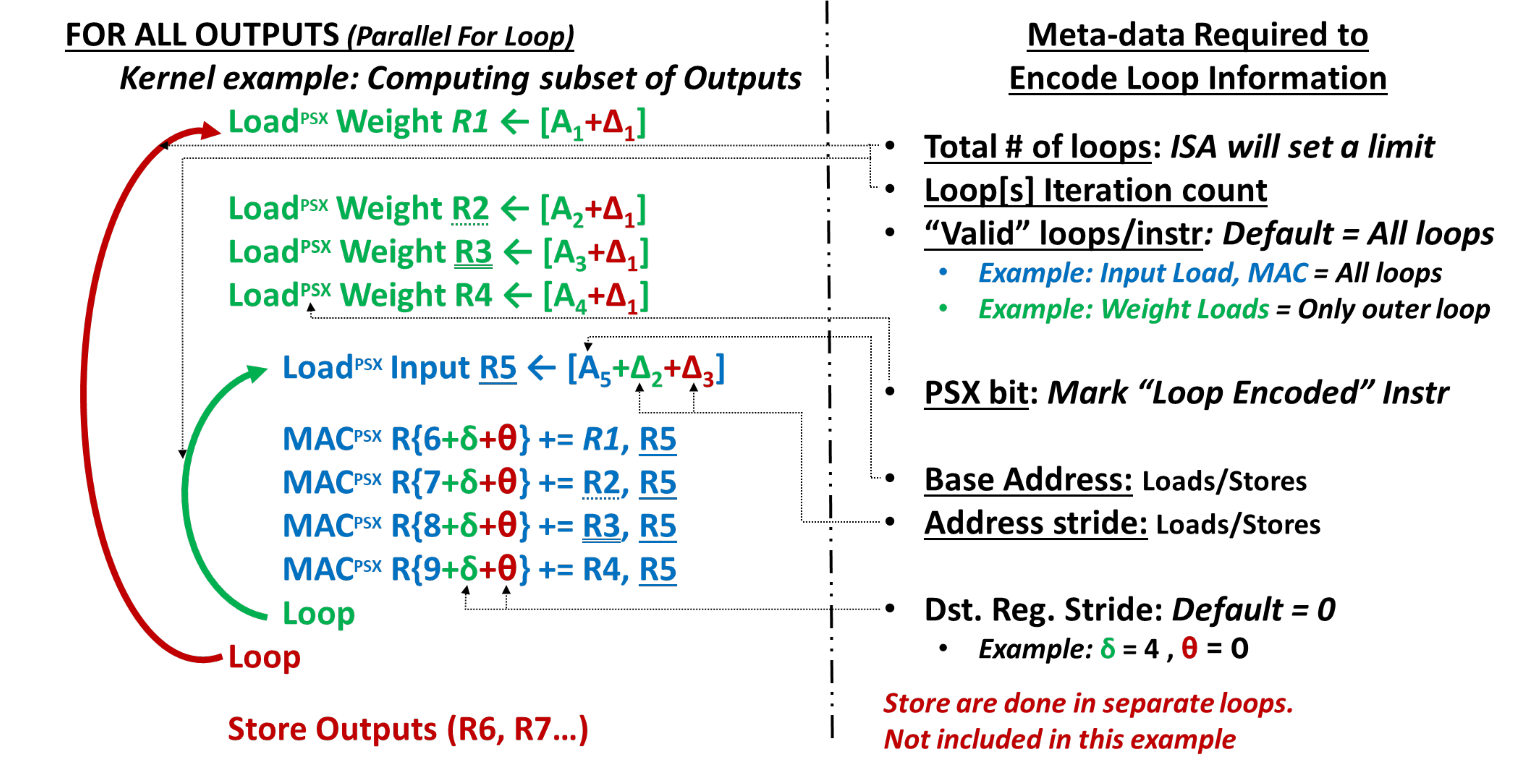}
  \caption{Requirements for encoding all loop information}
  \label{fig:psx_isa}
\end{figure}

A close examination of state-of-the-art MKL-DNN kernels implementing DNN
primitives shines light on opportunities to encode multiple loops of information
succinctly. \textit{Such optimizations would enable minimizing and bypassing the power-hungry
front-end stages of the CPU pipeline with unrolling and dispatch of work
proximal to the execution units}.

Depicted in Figure~\ref{fig:psx_isa}, is the meta-data information
each instruction requires to encode it's loop behavior in the kernel. First, we
need to know the number of loops and their iteration counts. The ISA can set a
limit on the maximum number of loops encoded, and we find that supporting four loops is
sufficient for all these kernels (capturing load, compute and store operations).
Each instruction needs to know the set of loops it resides within. In the
example, weight loads are only executed in the outer loop. Second, loads and
stores need a base address as well as an address stride for each loop it resides
in. These can be computed since DNN primitive implementation
employs structured data layouts to maximize cache port width and capacity.
Finally, data dependence is still through registers. Hence, we can require
stride values per loop for destination register ids as well. In the example,
iterations of the innermost loop reuses weights (being loaded in the outer loop) to
compute different output elements that need to be stored in different
registers. Here, the destination register id stride is determined by the number
of outputs computed in the innermost loop (4).

\begin{figure}[h]
	\centering
	\includegraphics[width=0.6\linewidth]{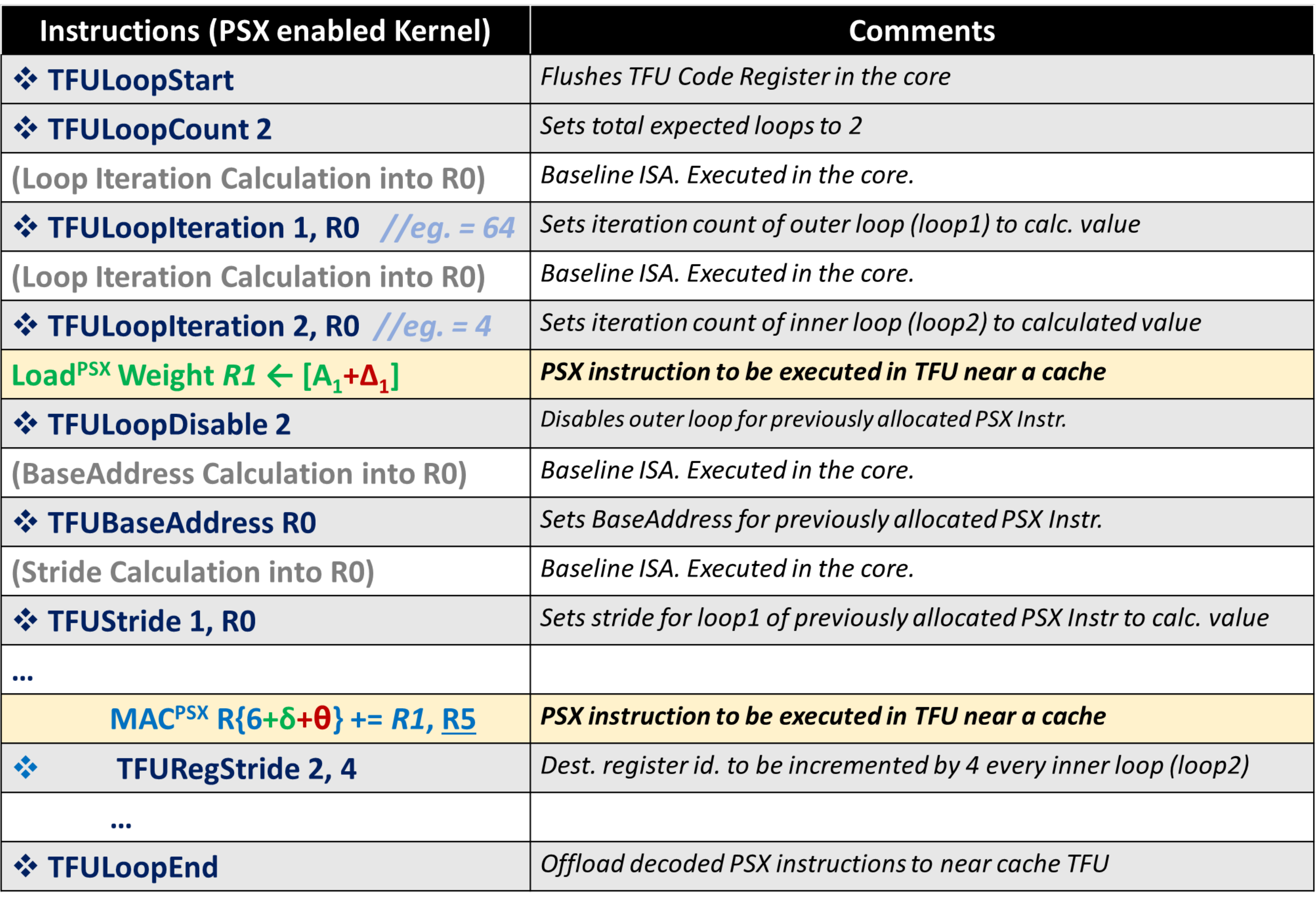}
	\caption{\proximusspace New PSX instructions and execution semantics}
	\label{fig:psx_flow}
\end{figure}

Figure~\ref{fig:psx_flow}, illustrates the new PSX instructions and their
semantics. Kernel instructions are tagged with a \texttt{PSX-bit} (denoting
near-cache TFU execution) and are decoded and allocated into new \texttt{TFU
Code Registers} in the core. Our examination of primitives across multiple DNN
models shows 32 registers to be sufficient. However, if a kernel has more than
32 instructions (and/or 4 loops) it would need to be split into smaller kernels
that fit within these constraints. New PSX instructions (\texttt{TFULoopCount},
\texttt{TFULoopIteration}, \texttt{TFULoopDisable}, \texttt{TFUBaseAddres},
\texttt{TFUStride}, and \texttt{TFURegStride}) populate their respective
meta-data loop information for the instructions tagged with PSX-bit. The meta-data
information can be calculated using regular ISA (similar to the way the baseline
kernels currently do it). The new \texttt{TFULoopStart} instruction flushes the
\texttt{TFU Code Registers} for new PSX-tagged instructions and the
\texttt{TFULoopEnd} instruction dispatches the \texttt{TFU Code Registers} to
the near-cache TFU for unrolled execution. We conservatively estimate each
\texttt{TFU Code Register} holds 8B of information (opcode, up to 3 registers (or
a register and base address),4 loop iteration counts (with a valid bit) and 4
address and register strides). {\em The entire offload takes 16 cycles (8B offload
bus width) and this time is amortized by the hundreds of cycles of
unrolled execution in the TFU.}

\subsubsection{Tensor Functional Units~(TFU)}

\begin{figure}[h]
  \centering
  \includegraphics[width=0.4\linewidth]{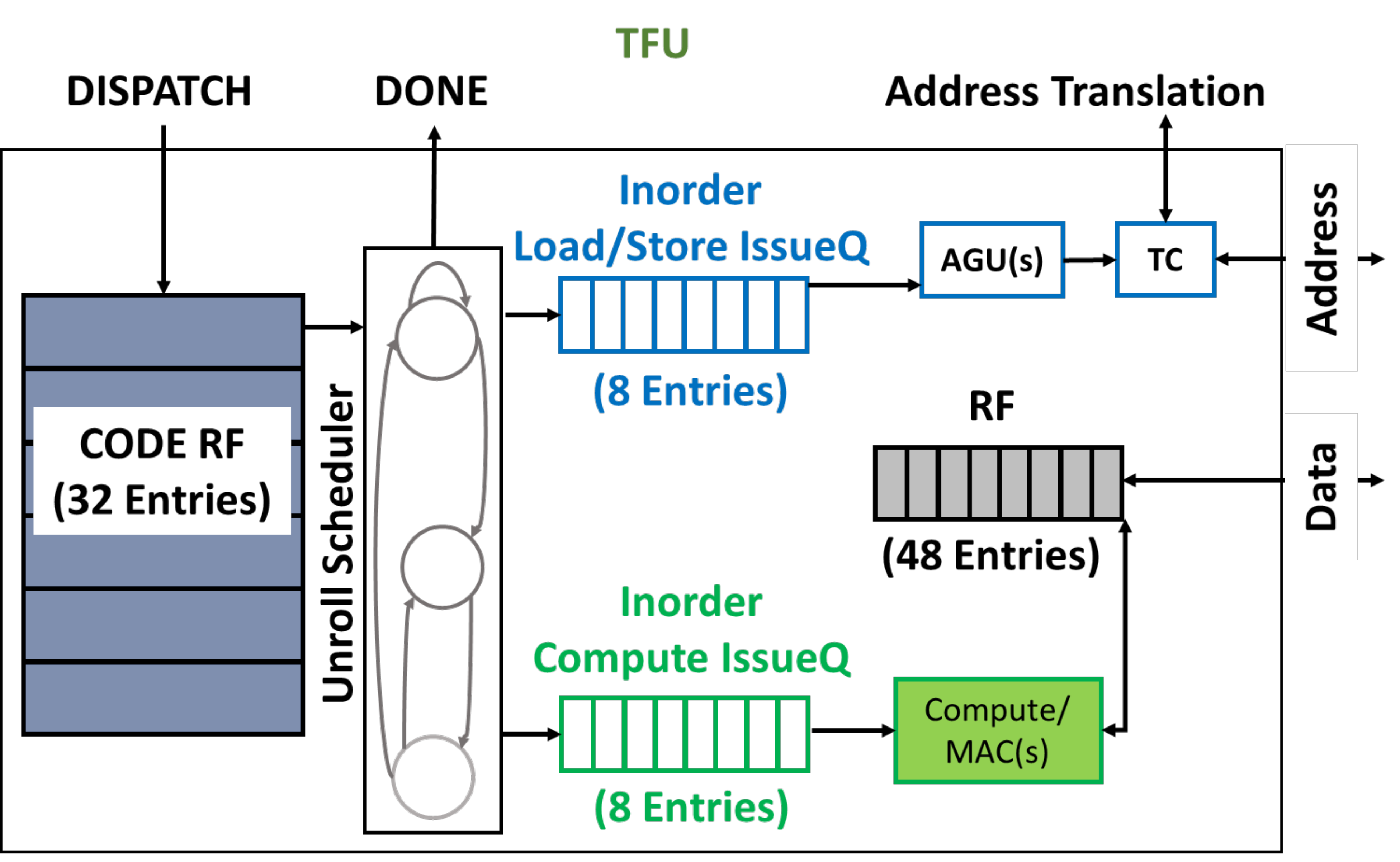}
  \caption{Schematic of the Tensor Functional Unit (TFU)}
  \label{fig:tfu}
\end{figure}

Figure~\ref{fig:tfu} depicts the Tensor Functional Units~(TFU), with 16
\texttt{TFU Code Registers}. A \textit{lean} ``Unrolling Scheduler" populates
two 8-entry \textit{in-order} \texttt{Issue Queues} - one for all compute
opcodes and another for loads and stores. The design simplifies scheduling
(\textit{lower power!}) and allows hoisting of loads over compute (to hide load
latency) while maintaining strict load/store ordering within the TFU. Loads and
stores directly access the cache each TFU is placed proximal to - bypassing any
inner levels. Snoops into inner cache levels, as required, are handled through
added coherency support to the cache~(Section~\ref{sec:coherency}). A small
Translation Cache assists in memory management~(Section~\ref{sec:VM}). 

Both kernel characterization and performance analysis show that a 48-entry
``deep" \texttt{TFU Data Register File} per TFU is sufficient - with no register
renaming (\textit{lower power again!}) required.
The loads/MAC-instr requirement of the workload and the near cache bandwidth
bounds the peak compute ``width" (the number of 64B MAC execution units) of the
TFU. We evaluate the performance, power and energy implications of different
compute widths in Section~\ref{sec:results}. TFU area analysis is included in Section
~\ref{sec:eval}.

\subsubsection{Leveraging SMT}

\begin{figure}[h]
  \centering
  \includegraphics[width=0.6\linewidth]{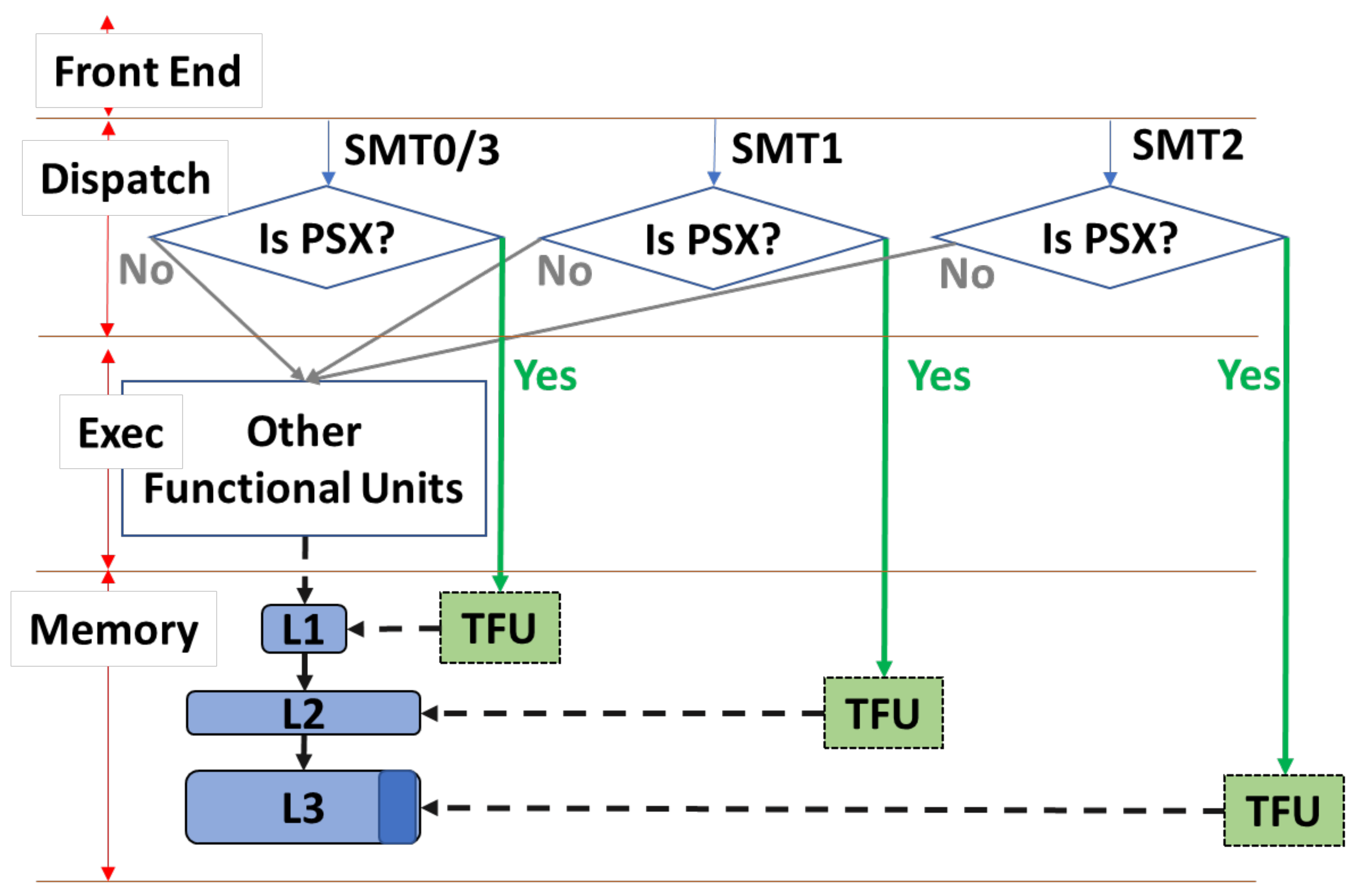}
  \caption{Leveraging SMT to bind each TFU to an OS-visible thread}
  \label{fig:smt}
\end{figure}

Modern server-class CPUs support 2-way (Intel, AMD) and 4-way SMT (IBM, Sun
SPARC). However, since all compute is shared across SMT threads, DNN frameworks
disable SMT or use only one thread for the primitives, relying on multi-core
compute scaling instead. With \proximus, each TFU is essentially a lean compute
engine directly accessing one cache level in the hierarchy. As shown in
Figure~\ref{fig:smt}, we leverage SMT to bind each TFU exclusively to
one of the logical SMT threads in the physical core. Therefore, each TFU is part
of a fully capable, OS-visible hardware context. \textit{DNN frameworks can
then distribute work across TFUs using existing threading runtimes}.  This
also enables fine grained control over which caches and TFUs to use for a
primitive(Section~\ref{sec:optimal_tfu}). PSX ISA, TFU design and SMT usage
allow \proximusspace to maintain the CPU memory model~(Section~\ref{sec:ordering}).

\subsection{\proximus: Micro-Architectural Support}
\label{sec:des_uarch}

\subsubsection{Virtual Memory}
\label{sec:VM}

The TFU AGUs compute the virtual address for loads and stores while caches in
modern CPUs are physically tagged. However, MKL-DNN optimizations use structured
and special layouts, customized to feed compute, with a high spatial locality of
tensor accesses~\cite{web:mem_format}. Through extensive characterization across
all primitives and models we find that a small 6-entry Translation Cache (TC),
holding recently observed virtual to physical mappings, can achieve a 90\%
hit-rate. Misses in the TC (10\%) can go through the existing TLBs and Page Walkers of
the local physical core for translations without adding any significant
bandwidth pressure on them. To ensure the TC entries are fully coherent, any TLB
invalidation or page swap invalidates all TC entries in all TFUs. 

\subsubsection{Distributed L3 caches}

L3 caches in modern CPUs are multi-bank structures shared across multiple-cores.
Any cache-block aligned address can reside in only one L3 bank. This presents a
challenge for TFU compute placed near each L3 bank. Multiple near-L3 TFUs will
likely need to access the same addresses (example: a weight element used to
compute multiple different output elements mapped to different TFUs in
convolution). Depending on the reuse out of the TFU RF, addresses may need to be
loaded multiple times. Traversing the L3 interconnect for every address not
available locally would cripple near-L3 TFU performance and add significant
extra data movement overhead. We leverage existing technologies like Intel's
CAT~\cite{web:intel_cat} or ARM's cache lockdown~\cite{web:arm_cache_lockdown}
to simply partition a portion of the set-associative cache (a small subset of
its total ways) in each L3 bank as a local cache for the attached TFU (with
added coherency support).  Section~\ref{sec:results} includes performance
sensitivity to the reserved local cache capacity for each near-L3 TFU.

\subsubsection{Coherency Support}
\label{sec:coherency}

Since the TFUs represent new loads, stores and compute near each level of
on-die cache, \proximusspace needs small additional tracking to maintain
overall cache coherence. L2 caches need an extra bit per cache-line to denote
whether L1 currently owns the cacheline, to ensure that it has ownership before
doing a store. Similarly the directory entries at L3 need an extra bit per
near-L3 TFU in their ``sharer/owner" vector structures. This denotes whether
the partitioned, local scratch-pad ways in that L3 bank have/own the
cache-line. Hence, \textit{\textbf{\proximusspace maintains the baseline CPU cache
coherence}}, generating appropriate snoops at L2 and L3 as required. 

\subsubsection{Memory Model (Ordering)}
\label{sec:ordering}

\textbf{\textit{\proximusspace maintains the Total Store Order~(TSO) memory
model in CPUs}}. Within a TFU, strict loads/store ordering is maintained. We add
a hardware fence in the core to \textit{prevent simultaneous TFU/non-TFU
execution within a thread}. A bulk offload of hundreds of cycles of work to the
TFU (through PSX-ISA) amortizes any performance cost of this serialization.
Non-TFU load/store ordering continues to be maintained by the core. Since TFUs
are on different SMT threads, we do not need to guarantee any ordering in
execution of loads and stores across TFUs.

\subsubsection{Handling Context Switches and Exceptions}

TFUs signal exceptions like existing functional/compute units in the physical
cores.  To support context switches on a thread, the core must also save/restore
the \texttt{TFU Code Registers} and \texttt{TFU Data Registers} for the TFU on
that thread as well as invalidate the local Translation Cache.

\subsection{\proximus: Programming Model Support} 
\label{sec:des_pg}

\subsubsection{Generating PSX Code}
High performance libraries already implement primitives using optimized code that
is JITed~\cite{doc:anatomy_paper}. The kernel instructions
must now be tagged with the \texttt{PSX-bit}. The JITer
is already used to compute various loop variables. We extend the JITer to add
\texttt{PSX-bit} tags to the new PSX instructions that populate the loop
variables. Note that the programmer need not be aware of the exact TFU and cache
level where these PSX instructions will eventually be executed. Optimizing
compilers to generate PSX code from native languages without JITing is possible
but we do not explore that direction in our work.

\subsubsection{Exposing Proximu$\$$ Capability at Each Cache Level}
Presence of TFUs in each cache level and their corresponding compute width is
exposed through the \texttt{cpuid} interface that is supported by all modern
processors.

\subsubsection{Optimal TFU selection for primitives}
\label{sec:optimal_tfu}
As we characterize and summarize in Table~\ref{table:tab_summary}, and with further analysis in
Section~\ref{sec:results}, each primitive has an optimal set of TFUs for
power-performance efficiency. We leverage the use of SMT (binding a TFU to a
logical thread) and use existing OpenMP APIs to set the affinity of a primitive to a
subset of cores. Specifically, we use \texttt{KMP\_SET\_AFFINITY} in the LLVM
OpenMP Runtime~\cite{web:llvm_openmp} to achieve this. DNN frameworks
(TensorFlow, Caffe etc) would do the same before invoking the MKL-DNN
implementation of a primitive (similar to currently only using one thread per
core).

\subsubsection{Distribution of work across TFUs}
Since caches across the multi-level hierarchy have different bandwidths, their
corresponding TFU will have different compute ``widths" as well. For compute
bound primitives like convolution, we need to divide work across TFUs
proportionate to their compute strength for optimal performance. With high cache
hit-rates and predictable performance, such a ``static" division of work is
sufficient for these workloads. Towards this, we introduce a simple new schedule
{\em kind} called {\tt static\_asymmetric} in the LLVM OpenMP runtime. For
example, for three TFUs with compute strengths in a 2:2:1 ratio, a static equal
division of work would result in unequal thread completion times with the third
(weakest) TFU determining final runtime. With a static asymmetric distribution,
in the same 2:2:1 ratio, all threads optimally complete at the same time.

\begin{table}[htbp]
  \centering
  \caption{Changes in software stack to support \proximus}
    \setlength\tabcolsep{2pt}
    \begin{tabular}{|p{13.0em}|p{15.0em}|} 
    \hline
    \multicolumn{1}{|c|}{\textbf{Level in Software Stack}} & \multicolumn{1}{c|}{\textbf{Proximu\$ Support}} \\\hline
    DNN Frameworks \newline{}(Tensorflow, Caffe, PyTorch) & Set \texttt{KMP\_THREAD\_AFFINITY} appropriately for each primitive \\\hline
    High Performance Library \newline{}(MKL-DNN) & Use PSX extensions for tensor load, store, compute instructions \\\hline
    Threading Runtime \newline{}(OpenMP) & Asymmetric static scheduling when appropriate (e.g. Convolution) \\\hline
    \end{tabular}%
  \label{table:sw_stack}%
\end{table}%

\subsubsection{Summary}
Table~\ref{table:sw_stack} summarizes the intercepts in the overall software
stack by \proximus, with \textbf{\textit{no changes to the current CPU programming model}}.

\section{Evaluation Methodology}
\label{sec:eval}
\textbf{Simulation Framework:} We use a modified version of
Sniper~\cite{doc:sniper} for our cycle-accurate multi-core simulations. We
model a state-of-the-art Intel 28 core datacenter processor
~\cite{web:intel_clx} but with 4-way SMT whose parameters
are listed in Table~\ref{table:sim_parameters}. This baseline supports a peak
128~(2*64) MACs/cycle/core of compute (similar to Intel
DL-Boost~\cite{rodriguez2018lower}) with per-core on-die cache bandwidth
including two 64B read ports at L1, two 64B read/write ports at L2 and one 64B
read/write port at L3. 

The simulation framework has been thoroughly validated against silicon by
executing all the evaluated MKL-DNN~\cite{web:mkldnn} primitives and verifying the performance
trends using a Cascade Lake system running Ubuntu 16.04 with MKL-DNN v1.0
library. CACTI~\cite{cacti},\cite{pcacti} and McPAT~\cite{mcpat} are used to
quantify cache and overall energy impact.  
\begin{table}[h]
    \centering
    \caption{Simulator Parameters}
    \scriptsize
    \begin{tabular}{|m{5em}||m{23em}|}
         \hline
         \textbf{28 Cores} & 2.6GHz, 4-way SMT, 320-entry ROB, \newline{}128
		 (2*64) MACs/cyc/core\\
         \hline
         \textbf{L1 cache} & Private, 32kB, 8-way set associative, LRU, 8-enrty MSHR, 2x 64B read ports, 1x 64B write ports, Data access=4 cycles, Tag lookup=1 cycle\\
         \hline
         \textbf{L2 cache} & Private, 1MB, 16-way set associative, LRU, 48-enrty MSHR, 2x 64B read/write ports , Data access=8 cycles, Tag lookup=2 cycle\\
         \hline
         \textbf{L3 cache} & Distributed, Non-inclusive, 1.375 MB/slice, 11-way set associative, RRIP, 48-enrty MSHR, 1x 64B read/write port per slice, Data access latency=10 cycles\\
         \hline
         \textbf{Tag directory} & MESIF, 10 cycle latency \\ 
         \hline
    \end{tabular}
    \label{table:sim_parameters}
\end{table}

\begin{table}[b]
    \centering
    \caption{Notation for \proximusspace configurations}
    \scriptsize
    \begin{tabular}{|c||c||c|}
    \hline
    Name & MACs/Cycle/Core & Details\\\hline\hline
         \textbf{P128} & 128 & same as M128\\\hline
         \textbf{P256} & 256 & 128@L1, 64@L2, 64@L3\\\hline
         \textbf{P320} & 320 & 128@L1, 128@L2, 64@L3\\\hline
         \textbf{P512} & 512 & 256@L1, 128@L2, 128@L3\\\hline
         \textbf{P640} & 640 & 256@L1, 256@L2, 128@L3\\\hline
    \end{tabular}
    \label{tab:notation}
\end{table}
\begin{table}[t]
    \centering
    \caption{Area Breakdown for TFU (in $mm^2$)}
    \scriptsize
    \begin{tabular}{|c|c|c|} \hline
    Registers & MACs &TC,Queues,Control\\\hline
    0.15 & 0.17 & 0.06\\\hline
    \multicolumn{3}{|c|}{\textbf{Total Bytes: 3184}}\\\hline
    \multicolumn{3}{|c|}{\textbf{Total Area: 0.38$mm^2$}}\\\hline
    \end{tabular}
    \label{tab:area}
\end{table}

\textbf{\proximus:} We model all of \proximusspace in Sniper allowing sweeps of peak compute at each
TFU at different cache levels. The last SMT thread is tied to L1 TFU but not
used. We use prefix ``\textbf{P}'' (for \proximusspace) or
``\textbf{M}'' (for traditional monolithic core) followed by a number which indicates the
peak number of MACs/cycle/core that the configuration has. Further, the notation
also attaches an explicit distribution of compute resources across the cache
levels as indicated in Table~\ref{tab:notation}. \textbf{Mxxx} configurations
have MAC units that support \textbf{xxx} MACs/cycle/core.
Table~\ref{tab:notation} specifies the hardware resources and the distribution for the \proximusspace
configurations.

\textbf{Area Requirements of the \proximus:} We synthesize a Verilog
implementation of a TFU instance that is capable of 256 (4*64) MACs/cycle/TFU to
support the P640 configuration.  Synthesis is done using TSMC $28 nm$ library,
setting a target clock of 1 GHz, using Synopsys Design Compiler
~\cite{web:synopsys_dc}. The total area per TFU is $0.38 mm^2$ and the detailed breakup area is given in Table~\ref{tab:area}. 
An additional 2KB/core of storage is required for new core-valid bits in L2 and
L3 for full coherence. Total area overhead is the sum of this coherence
storage and the area required for three TFUs/core. Projecting the total
overhead to the $14nm$ technology~\cite{scaling_vlsi_journal},~\cite{web:bohr} in which the Intel Xeon server chips are
built, \textit{we conservatively estimate the overall overhead due to \proximusspace to
be a mere 2.63\% of a single Xeon core}. From observing
die-plots(\cite{doc:dieplot},~\cite{web:server_die_area}), \proximusspace takes
35\% less than the AVX-512 unit already present in the cores, while delivering a
higher peak MACs/cycle/core cumulative compute.

\textbf{Workloads and Software Stack:}  We evaluate six DNN topologies
\textit{end-to-end} including ResNet-50~\cite{doc:resnet50},
DenseNet-169~\cite{doc:densenet}, MobileNet~\cite{mobilenet},
ResNext-101~\cite{resnext101}, Transformer~\cite{transformer} and
TwoStream~\cite{twostream}. Of these, Transformer comprises solely of
inner-product layers while the others have mostly convolution layers.
We use the
latest open source MLK-DNN v1.0 and the Intel C++ compiler 19.0 to ensure we are
evaluating state-of-the-art software implementations.
Note that these are full system (including DRAM), multi-core evaluations. Since we study int8
inference, most model weights fit in caches. Coupled with high reuse, overall
impact on performance and power from DRAM is very low.
These workloads are
parallelized using the OpenMP multi-threading framework using our new static
asymmetric scheduling.

\section{Results}
\label{sec:results}

We will first present performance and data movement impact of \proximusspace
near-cache compute for DNN primitives in the ResNet-50 and Transformer models.
We then do a detailed power, performance and energy study for \proximusspace on
ResNet-50 and Transformer. This is followed by results for all six DNN topologies
evaluated. We then summarize the performance and performance/watt goodness of
\proximusspace and conclude by comparing it with available MLPerf data for GPUs
and DSAs.

\begin{figure}[ht]
  \centering
  \includegraphics[width=0.7\linewidth]{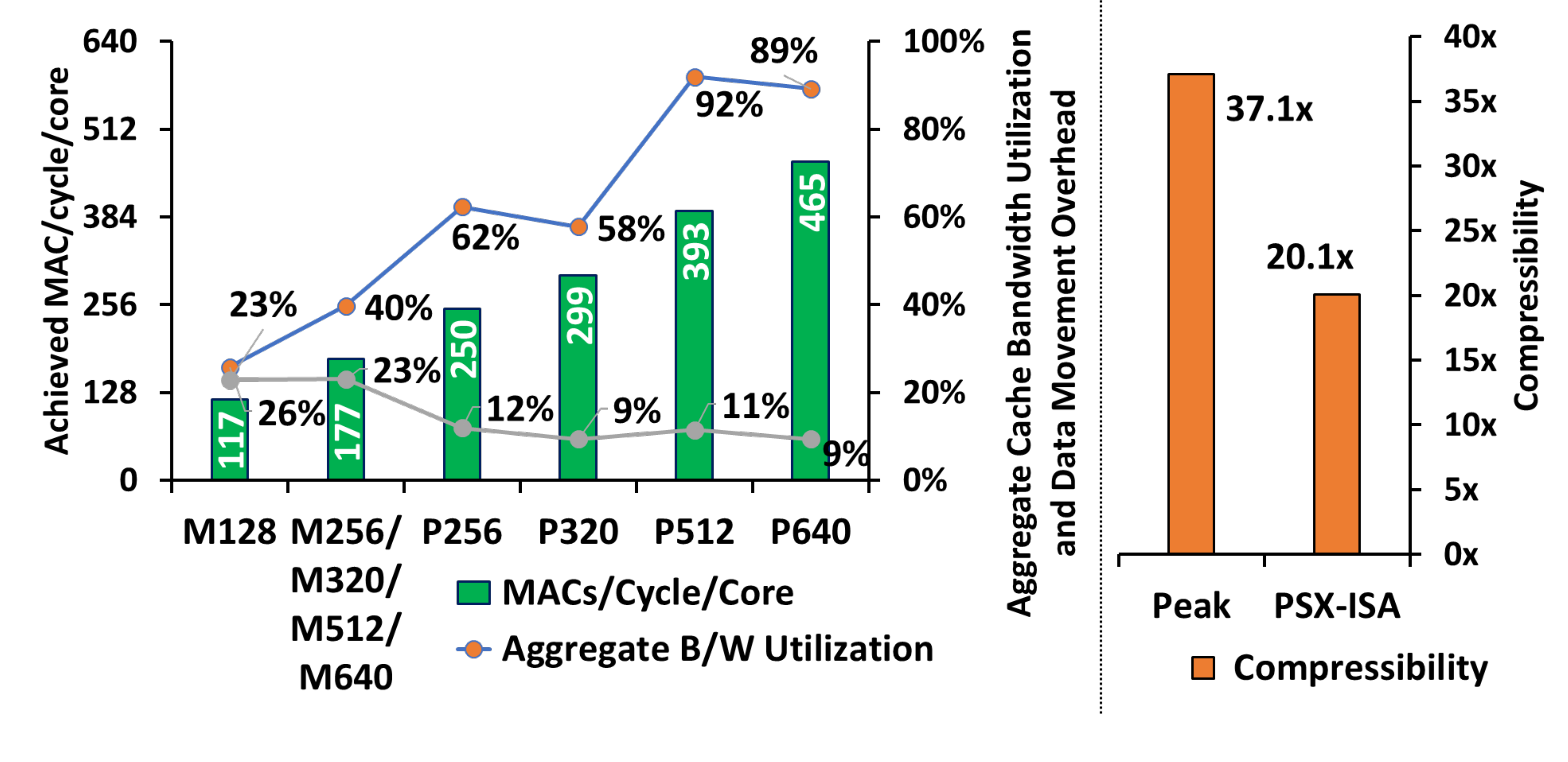}
  \caption{Impact of \proximusspace on ResNet50 Convolutional Layers}
  \label{fig:conv_perf_bw_dm}
\end{figure}

\subsection{Convolution}
Figure~\ref{fig:conv_perf_bw_dm} shows the impact of \proximusspace on ResNet50
convolutional layers and reveals multiple interesting observations. Traditional
scaling of compute plateaus at an average 180 MACs/cycle/core from M256
onwards, since L1 bandwidth saturates, despite using only 40\% of total
available on-die bandwidth. {\em However, \proximusspace scales well in performance at all
points, achieving between 2x to 3.94x performance over baseline, with up
to 90\% cumulative bandwidth utilization in the higher peak compute points}.
\proximusspace P256 also has 41\% higher performance than M256, achieving near
peak performance. M256 needs peak bandwidth (100\% hit-rate) from L1 to achieve
performance. In contrast, the P256 config doesn't require peak bandwidth from
any of its caches and is hence able to get higher performance. 

\textit{PSX-ISA achieves an average 20$\times$ reduction in dynamic
instructions executed in the legacy OOO pipeline}, which will translate to power
savings. Peak compressibility is actually 37$\times$, with the new
PSX instructions to populate loop meta-data accounting for the difference. Finally, \proximusspace reduces data movement
overheads from 20\% down to 10\%, mainly going down at the L1-L2 interface.

\begin{figure}[h]
  \centering
  \includegraphics[width=0.7\linewidth]{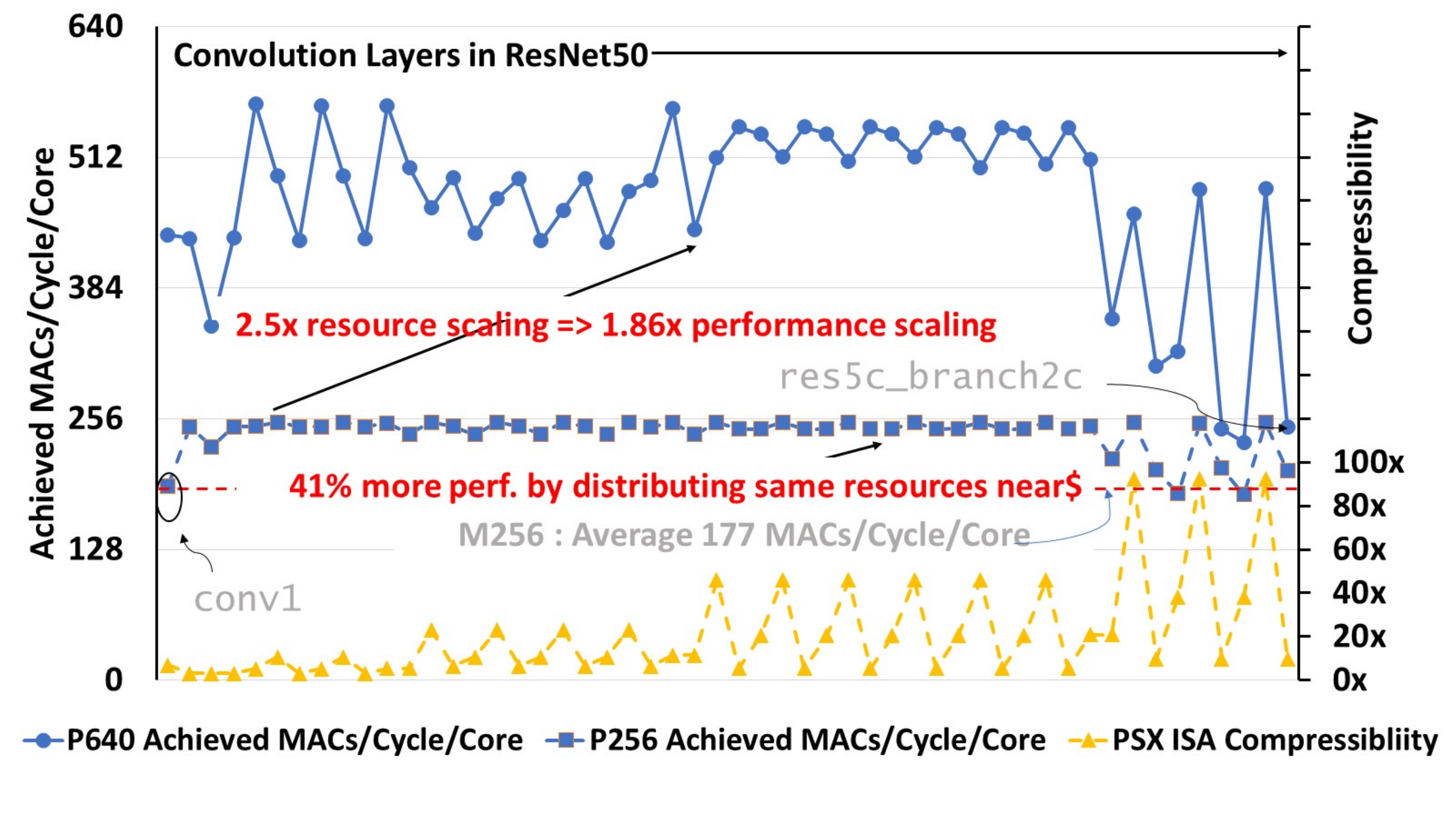}
  \caption{Per ResNet-50 convolution layer \proximusspace performance 
  for P256 and P640 configurations and Compressibility using PSX ISA}
  \label{fig:conv_perf_bw_layer}
\end{figure}

Figure~\ref{fig:conv_perf_bw_layer} shows per-layer performance and
compressibility for the 53 convolution layers in ResNet-50 for the P256 and P640
configurations. Note that the last few convolutional layers of ResNet-50,
including the \texttt{res5c\_branch2c} layer, achieve lower performance with
\proximus. These layers have lower total Ops/Byte with the near-L3 TFUs see low
hit-rates in their 256KB local partitioned cache (2 out of 11 ways), with high
cross-L3 traffic. Increasing the reserved near-L3 ways from 2 to 8 ways,
improves performance for these layers by 40\%-60\%. Compressibility increases
with increase in the input channel dimension (due to higher accumulation
required per output) and is generally lower for 1x1 kernels compared to 3x3
kernels (due to lower input reuse).

\subsection{Inner Product}
\begin{figure}[h]
  \centering
  \includegraphics[width=0.7\linewidth]{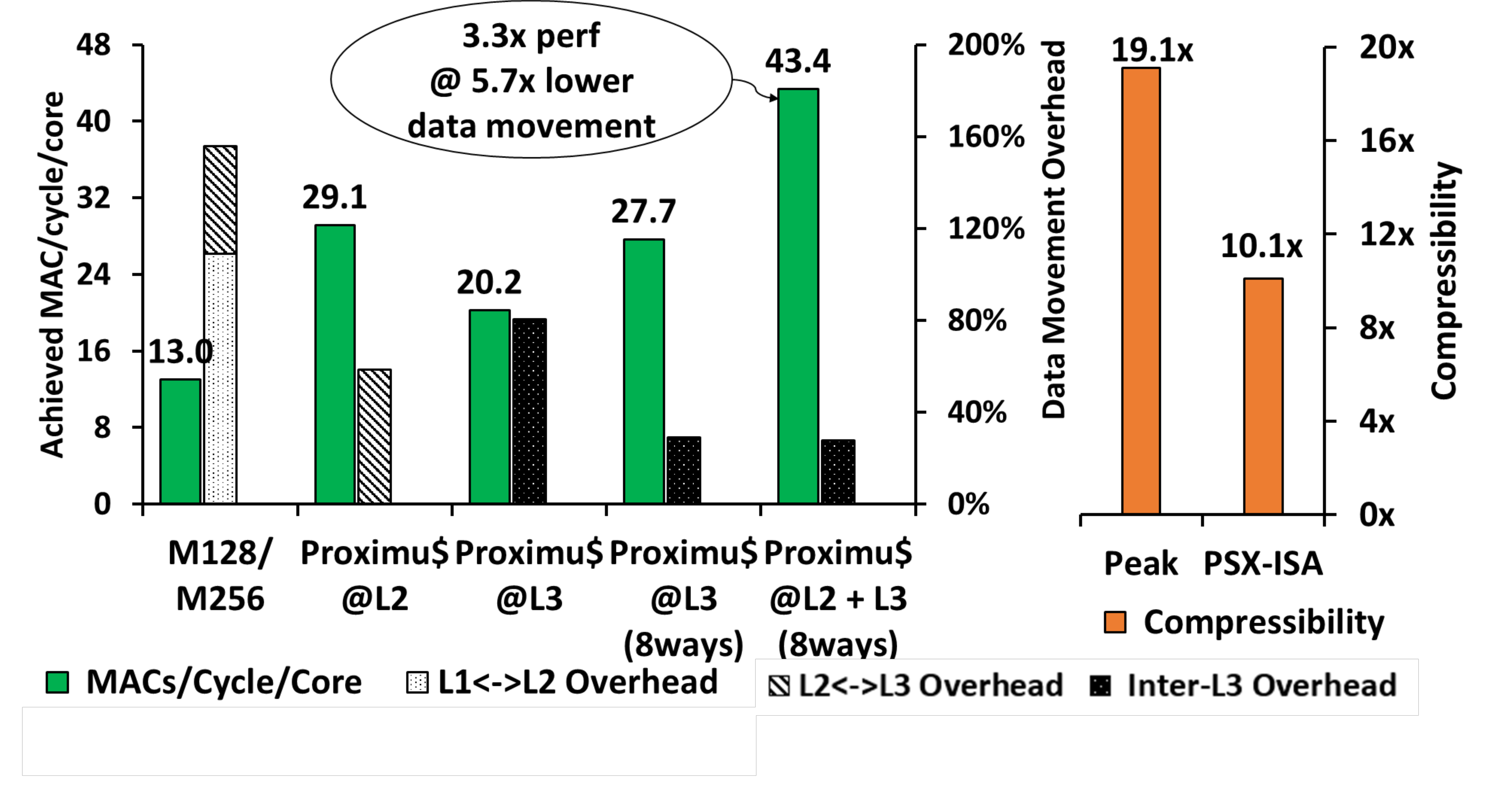}
  \caption{Impact of \proximusspace on Transformer Inner-Product Layers}
  \label{fig:fc_perf_dm}
\end{figure}
Figure~\ref{fig:fc_perf_dm} plots the average impact of \proximusspace
configurations on performance, data-movement for the 106 Transformer
inner-product layers. Since these are low Ops/Byte bandwidth bound primitives,
we only use the P256 \proximusspace configuration but schedule threads
selectively at different cache levels. By merely executing the inner-product
directly near the large 1MB L2, with better hit-rate and therefore bandwidth
delivery, \proximusspace achieves 2.2$\times$ more performance and a 2.6$\times$
reduction in data movement overheads. All L1 traffic has been eliminated.
Execution only near-L3, with a 2-way 256KB local cache, also reduces data
movement. However, as we can now expect, provisioning more capacity to the
near-L3 compute increases performance to match \proximusspace near-L2. Executing
the primitive at both L2 and L3 improves performance by a huge 3.3$\times$ over
what can be achieved by the baseline.  This is achieved with no increase in
on-die cache bandwidth while simultaneously reducing data movement overhead by
5.6$\times$. Finally, the new PSX-ISA achieves 10$\times$ compression, which
will directly translate to power and energy savings.

\subsection{Pooling and Concat Layers}
For brevity, we summarize the main observations from evaluations of the pooling and concat
primitives. Executing the res5c ResNet-50 pooling layer solely near L3 reduces data
movement overhead by 95\% (103\% down to 8\%). Similarly, Concat layers in
DenseNet-169 see an average 150\% data movement overhead, which is brought down
by 70\%-95\% (down to 25\%-5\% overhead) via near L2 or near L3 execution.

\subsection{Detailed Energy Analysis}
\begin{figure}[h]
	\centering
	\includegraphics[width=0.6\linewidth]{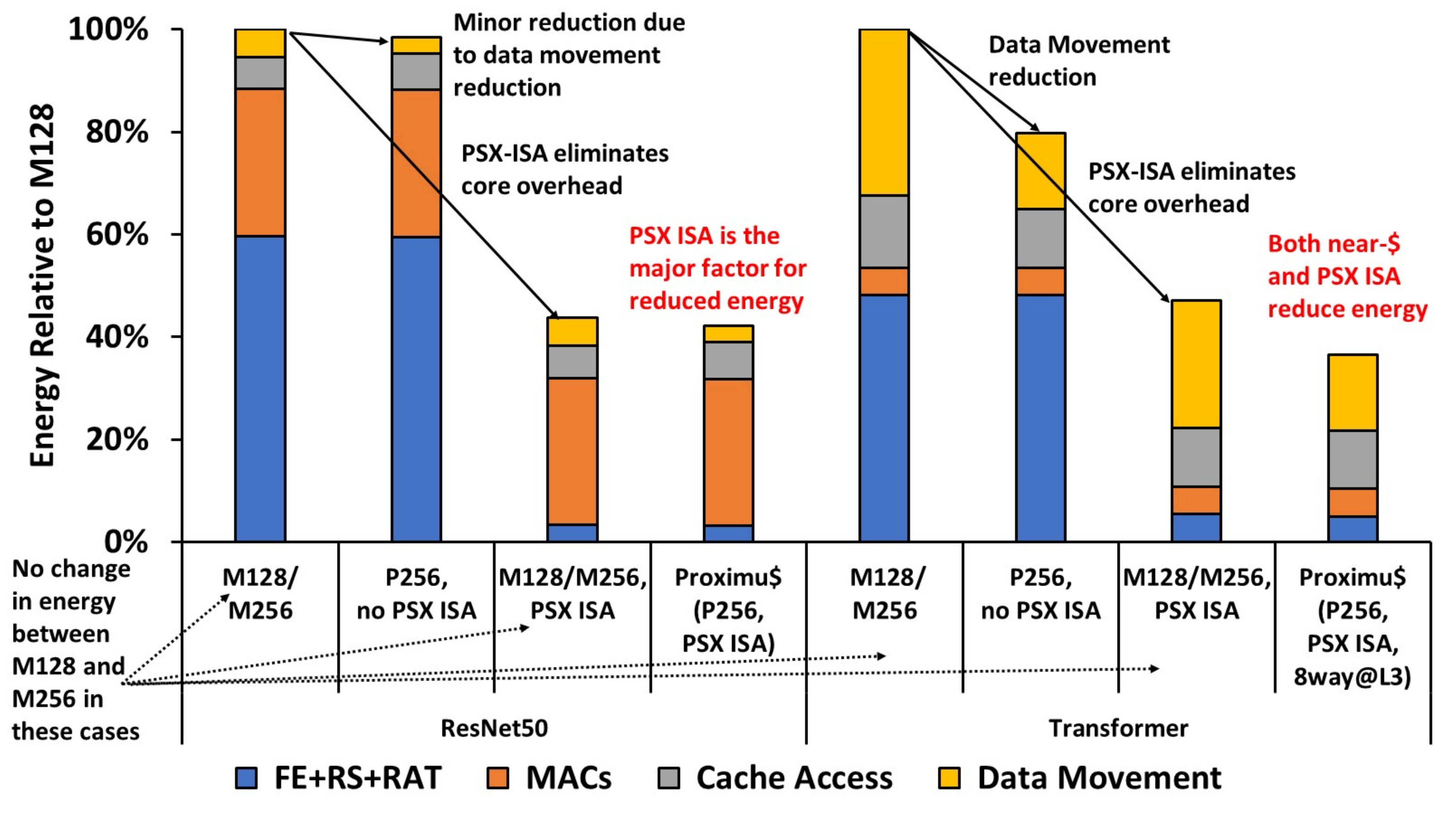}
	\caption{Stackup of Energy Consumption}
	\label{fig:energy_stackup2}
\end{figure}

Figure~\ref{fig:energy_stackup2} represents a detailed energy analysis of the
ResNet-50 convolutional layers and Transformer inner-product layers on the
baseline M128 and P256 configurations. All the energy components are shown
relative to the total energy cost of running on a M128 configuration. We
deconstruct the impact of the near-cache and PSX ISA components of the
\proximusspace proposal separately and when put together.

In ResNet50, the FE and OOO stages of the CPU pipeline dominate overall energy
(60\%).  M256, even while having twice the resources as M128, is iso-energy with
M128 as long as near-cache capabilities are not included to them. Adding
near-cache compute reduces inter cache data movement costs between L1 and L2 but
also increase direct accesses to the larger (costlier in power) L2/L3 caches.
Therefore, near-cache compute ends up iso-energy with baseline.  \textbf{Third,
due to high Ops/Byte and high reuse, the PSX ISA proposal achieves an average
20$\times$ compression, which translates to a 17$\times$ reduction in energy
from the FE and OOO stages of the pipeline}. \proximusspace P256 configuration
therefore operates at 42\% of baseline energy (translating to a 13\%
\textit{decrease} in power, along with a 2x \textit{increase} in performance).

\begin{figure}[h]
  \centering
    \begin{minipage}{0.5\textwidth}
    \centering
    \includegraphics[width=\linewidth]{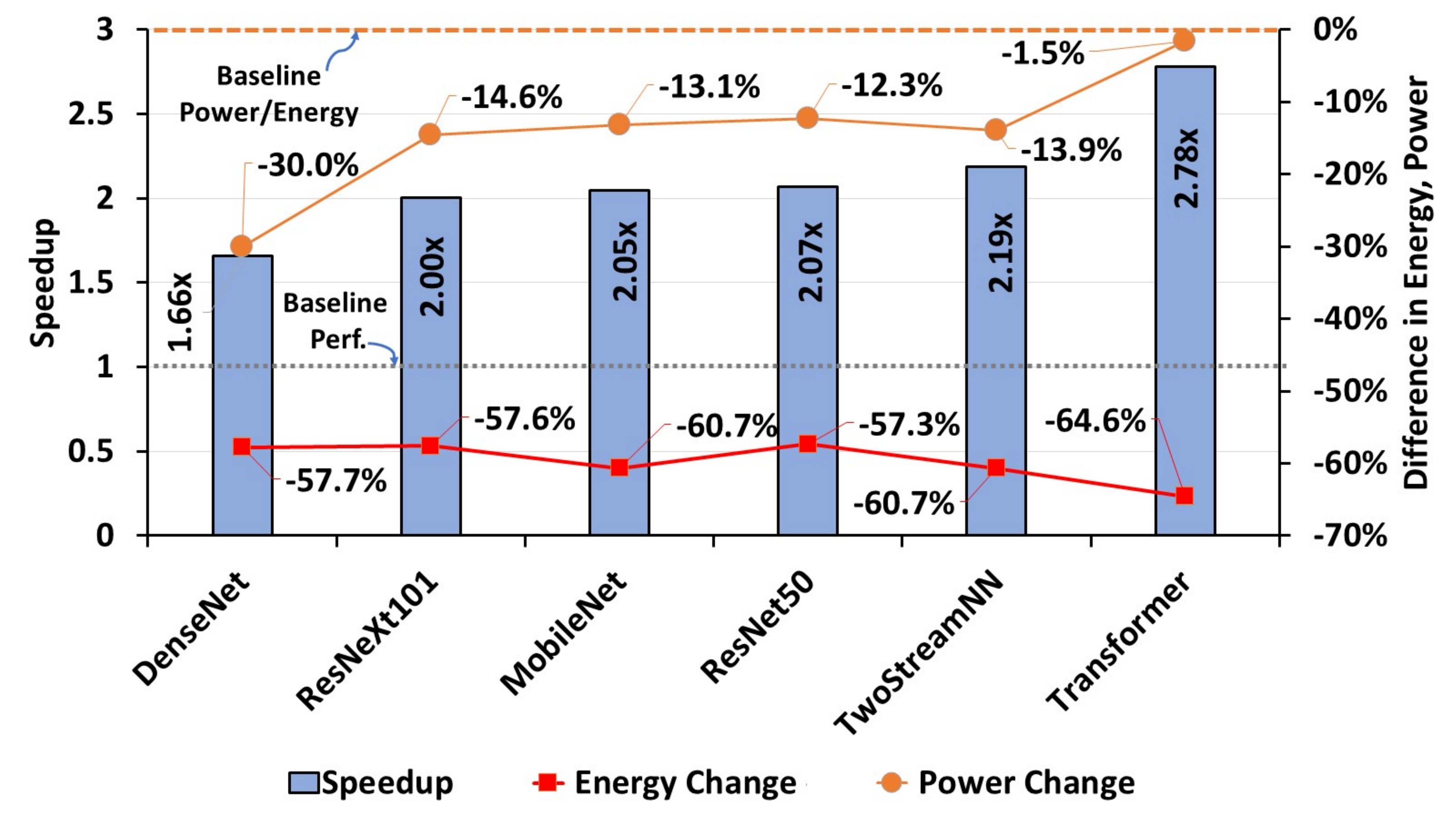}
    \captionof{figure}{Overall perf. improvement, energy reduction and power difference for six DNN topologies using P256 relative to M128}
  \label{fig:overall6top}
 \end{minipage}
    \begin{minipage}{0.5\textwidth}
  \centering
    \includegraphics[width=\linewidth]{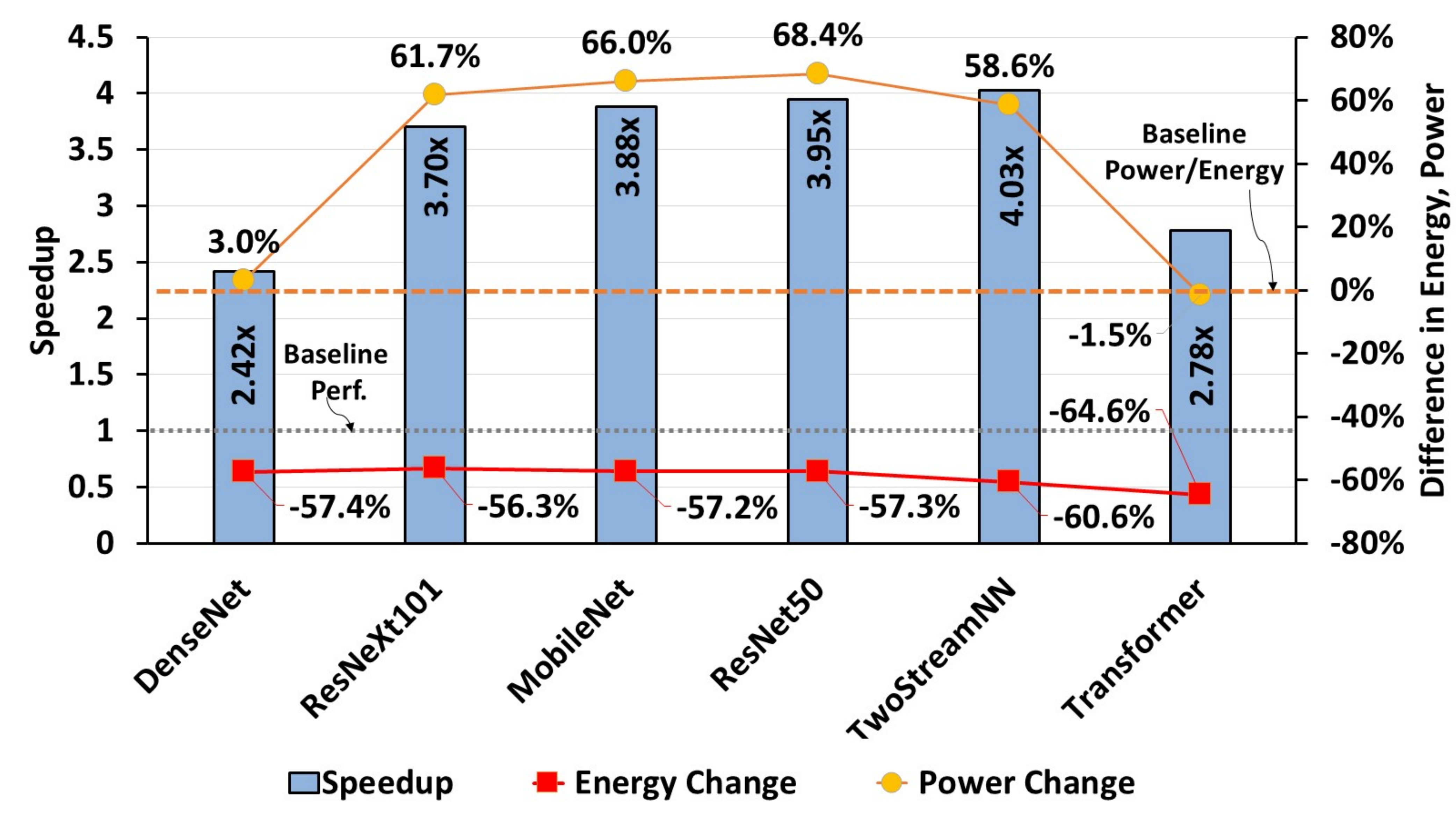}
    \captionof{figure}{Overall perf. improvement, energy reduction and power difference for six DNN topologies using P640 relative to M128}
  \label{fig:overall6_p442_m128}
\end{minipage}
\end{figure}

The bandwidth bound, low Ops/Byte inner-product layers in Transformer present a
different story: data movement reduction (primarily by eliminating L1) brings a
18.6\% reduction in energy. Despite lower data reuse compared to convolution,
the PSX ISA proposal achieves a 10$\times$ compression, bypassing the FE and OOO
pipeline stages and providing a 42.8\% energy reduction. Put together, we see a
61.5\% reduction in overall energy consumption (at 1.5\%
lower in power due to 2.77$\times$ better performance).

\subsection{Overall Performance, Power and Energy}

Figure~\ref{fig:overall6top} and Figure~\ref{fig:overall6_p442_m128} show
performance, energy and power impact of two \proximusspace configs (P256 and
P640 respectively) over six DNN-inference topologies. Inner-product heavy Transformer is
bandwidth bound. It achieves 2.78$\times$ better performance at iso-power (65\%
lower energy) for both \proximusspace configs. The rest of the DNN models
have mostly convolution layers. With the exception of DenseNet-169, these
topologies see around 2$\times$ performance with 2$\times$ compute (P256 vs
M128), at 12\%-14\% lower power (60\% lower energy). This increases to around
3.95$\times$ higher performance with 5$\times$ more compute (P640 vs M128) at 65\%
higher power (and 60\% lower energy). DenseNet-169 has a number of Concat layers
which take nearly 20\% of total runtime. \proximusspace reduces power for this
data shuffling primitive but doesn't impact performance. This results in slightly
lower performance improvement for DenseNet-169 with \proximus, but at slightly
better power compared to the other convolution heavy DNN topologies.

\begin{figure}[h]
  \centering
  \includegraphics[width=0.6\linewidth]{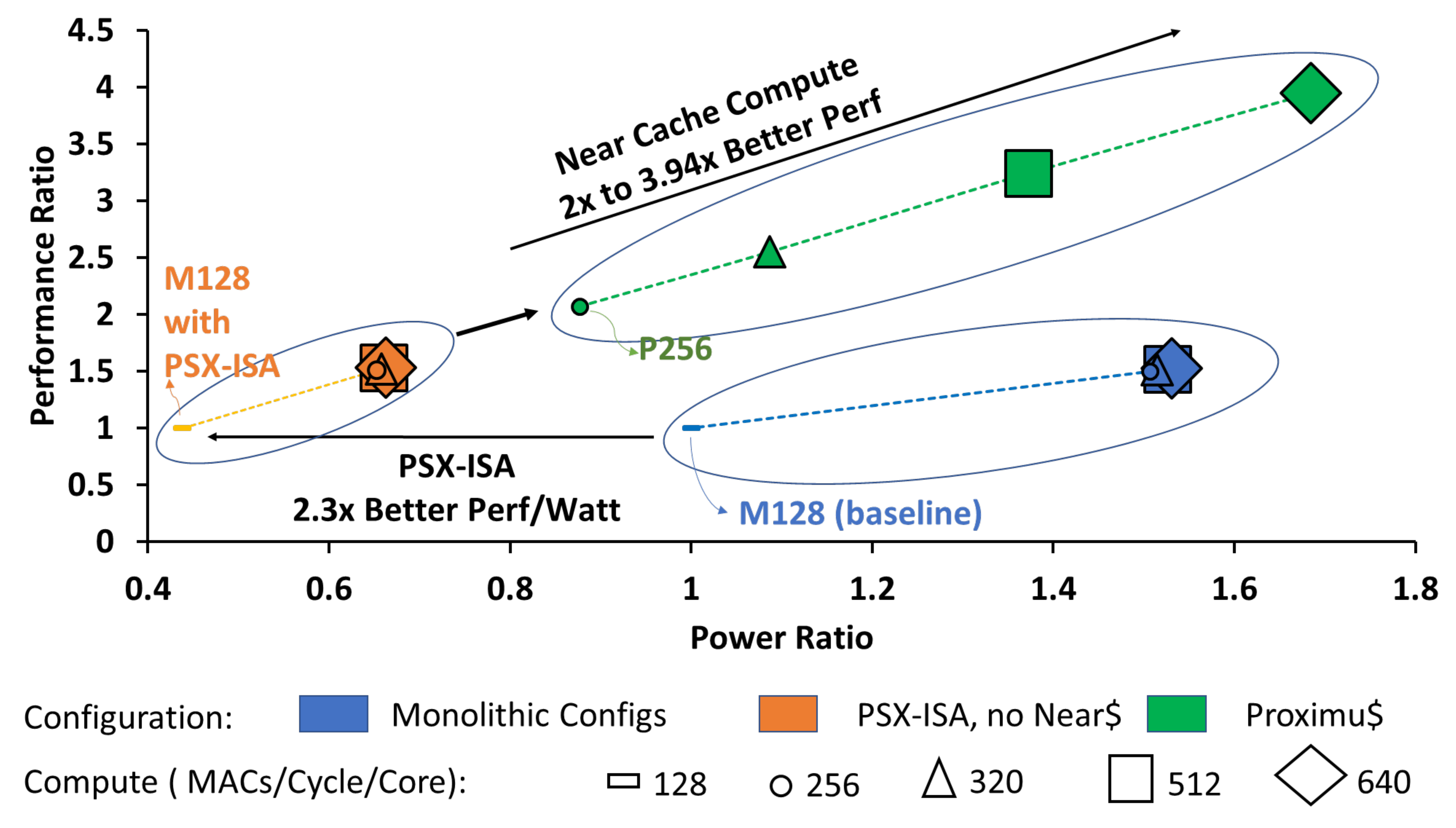}
  \caption{ResNet-50: Overall performance and power summary}
  \label{fig:ppw_resnet}
\end{figure}

\subsection{\proximusspace: Performance and Performance/Watt}
\label{sec:ppw}

Figure~\ref{fig:ppw_resnet} succinctly summarizes the goodness of \proximusspace
on ResNet-50. The PSX-ISA improves performance/watt by 2.3$\times$. Furthermore,
by leverage near-cache compute, performance can scale by 2$\times$ to
3.94$\times$ depending on the available TDP (13\% lower power to 68\% higher
power). Similarly, for inner-product heavy topologies like Transformer,
\proximusspace achieves a 1.8$\times$ increase in \texttt{inner-product}
performance/watt with 2.8$\times$ improvement performance. These gains are
achieved with minimal additional hardware, no increase in cache capacity or
bandwidth and no change in the CPU programming and memory models.

\subsection{\proximusspace comparison to DSAs and GPUs}
\begin{figure}[h]
  \centering
  \includegraphics[width=0.6\linewidth]{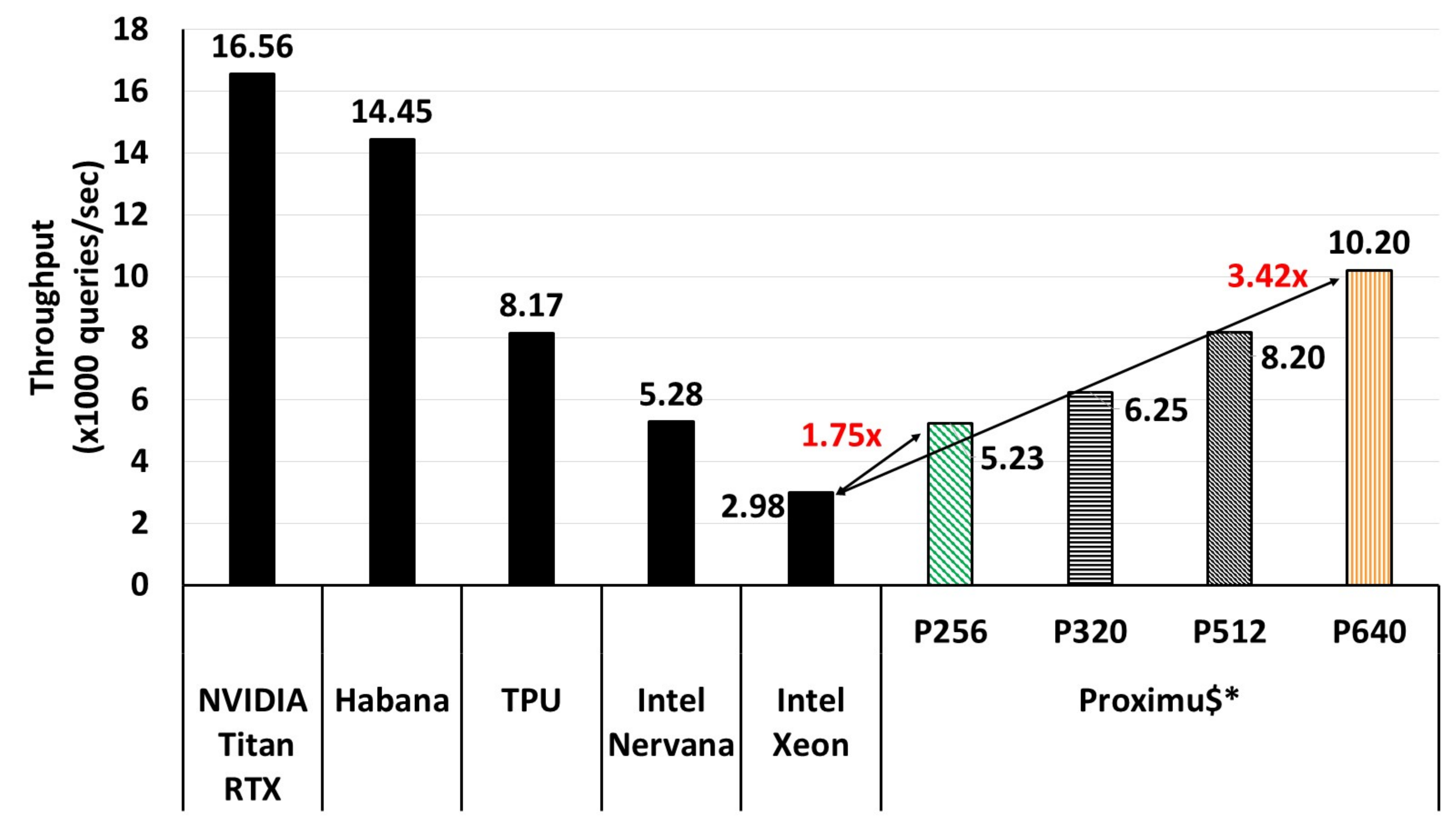}
  \caption{MLPerf Inference Throughput of GPUs, Accelerators, and CPUs compared
  with Projected Throughput$^{*}$ of \proximusspace}
  \label{fig:mlperf}
 \end{figure}

We compare \proximusspace with other platforms that are used for DNN inference
by using state-of-the-art, vendor-provided, publicly available throughput
results on ResNet-50 v1.5 from MLPerf Inference
results~\cite{web:mlperf},~\cite{doc:mlperf_paper}. Figure~\ref{fig:mlperf}
shows the number of queries processed by a variety of accelerators (Habana
Labs~\cite{web:habana} , Intel Nervana~\cite{web:nervana}, TPU
v3~\cite{doc:tpu}), GPUs(using a Titan RTX T496X2 card), and Cascade Lake based
Intel Xeon Platinum 9200 processor (similar to our CPU baseline). We normalize
the data to per node (for accelerators and GPUs) or per socket (for CPUs) and
use the best results for each platform. 

\proximusspace can achieve comparable (or even somewhat better) performance to
state-of-the-art DSAs (Intel Nervana, Google TPU v3). We don't make any
performance/watt comparisons due to the absence of these metrics in MLPerf.
However, we have shown significant (1.8$\times$-2.3$\times$) improvement in CPU
performance/watt with \proximus. \textit{\proximusspace requires significantly
lower area than the DSAs since it reuses existing cache, memory and scheduling
resources in the CPU}.

\subsubsection{Sensitivity to cache bandwidth}

\begin{figure}[h!]
  \centering
  \includegraphics[width=0.6\linewidth]{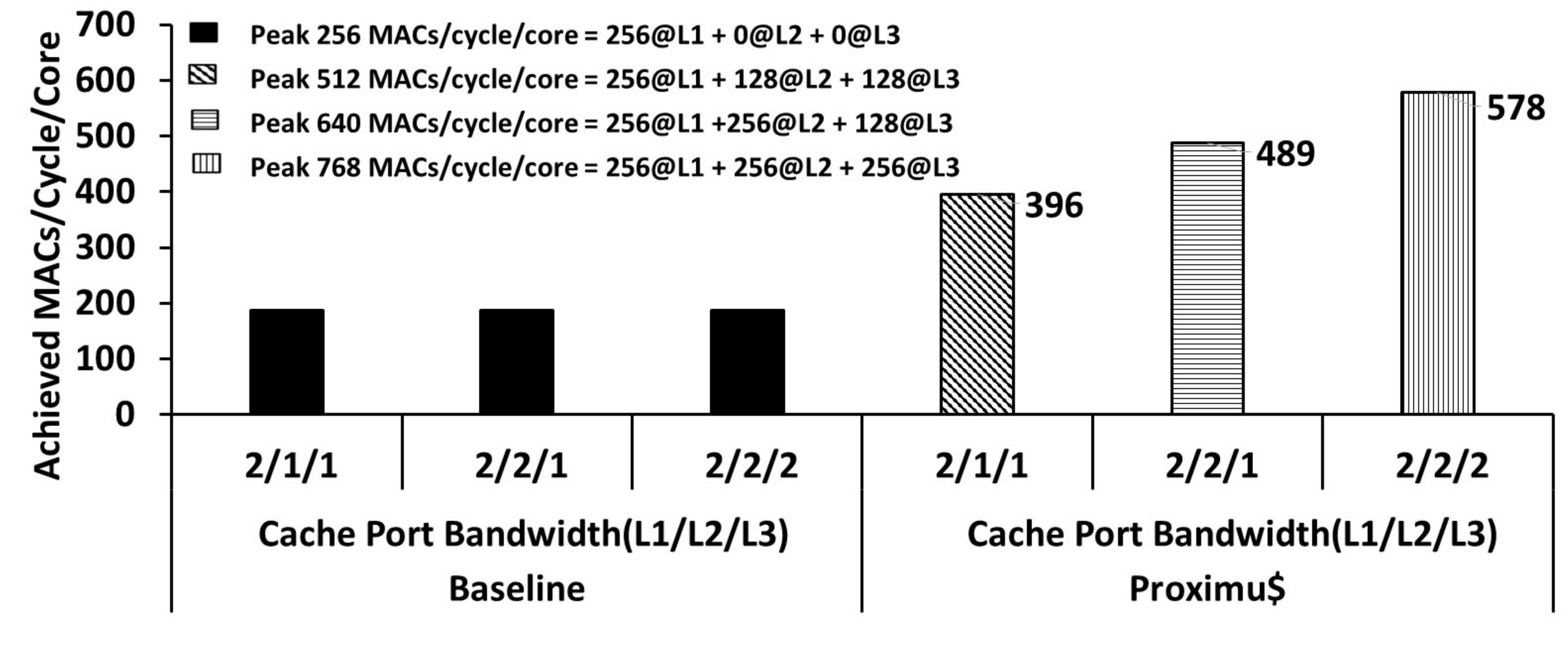}
  \caption{Sensitivity of \proximusspace to cache bandwidth scaling}
  \label{fig:bw_sensitivity}
\end{figure}

We further study the effectiveness of using \proximusspace in systems that have
different bandwidth configurations at L2 and L3 (reducing L2 bandwidth to a
single 64B port (2/1/1) or increasing per-slice L3 bandwidth to two
64B ports (2/2/2)). We follow our methodology of sizing compute
proportional to the associated cache bandwidth: for example, reducing to 128 MACs/cycle/core
near-L2 when it has a single read/write port (2/1/1). As shown in
Figure~\ref{fig:bw_sensitivity}, while the baseline performance plateaus beyond
the peak 256 MACs/cycle/core point, \proximusspace enabled systems continue to
scale performance. \textit{\textbf{\proximusspace achieves around 75\% compute
efficiency at every compute and bandwidth configuration}}.

\section{\proximusspace on low-power, edge CPUs}

We have verified the performance/power benefit of \proximusspace across a range
of compute widths, caches sizes and bandwidths; including lower compute (16/32
MAC/cycle/core) and cache bandwidths typical to lower power edge CPUs. These
systems can pose two additional challenges. First, they typically have shallower
cache hierarchies, often with multiple cores sharing an L2 cache. With
\proximus, we would still have a TFU per core at the L2 cache, but with lower
compute strength. The total compute strength across all TFU near a shared-L2
should be proportional to the L2 bandwidth. Second, these cores typically do not
have SMT capabilities. This essentially means a core would simultaneously
schedule to all its TFUs, breaking support for TSO memory ordering. Prior
studies~\cite{web:kunle-isca-18-keynote} have shown that relaxed consistency is
fine for ML workloads. For low power and low cost edge SOCs, where budget,
latency and battery life requirements reduce the ROI of adding specialized DSA
hardware with driver-based offload, we believe \proximusspace represents
an excellent CPU solution on all cost, performance and energy metrics. 

\section{Related Work}

Over the past few years, there have been numerous works targeting moving
processing to in or near memories like
DRAM(\kern-0.5em\cite{doc:rowclone},\kern-0.5em\cite{doc:ambit},\kern-0.5em\cite{doc:dracc}),
3D-stacked HBM\kern-0.5em\cite{web:hbm} or
HMC\kern-0.5em\cite{doc:hmc}(\kern-0.5em\cite{ahn2016scalable},
\kern-0.5em\cite{zhang2018graphp},\kern-0.5em\cite{dai2018graphh},\kern-0.5em\cite{krause2019nemesys},\kern-0.5em\cite{doc:impica},\kern-0.5em\cite{doc:google_consumer})
and even SSDs(\kern-0.5em\cite{doc:graphssd}).  Compute
Cache\kern-0.5em\cite{doc:computecache}, Neural
Cache\kern-0.5em\cite{doc:neuralcache}, and Duality
Cache\kern-0.5em\cite{doc:dualitycache} are all recent works that propose
converting the CPU on-die caches into compute units capable of bit-serial
/bit-parallel operations \textbf{in} the SRAM sub-arrays. While these works show
hugely impressive gains, they involve changes to highly optimized SRAM
sub-arrays, and can degrade or complicate signal integrity, density, floorplan.
\proximusspace takes a \textit{light-weight, compute-\textbf{near}-cache approach,
reusing existing micro-architectural interfaces} and extracting large
improvements in performance, power and area utilization, all with no changes to
the existing programming model.

\section{Summary}
\label{sec:summary}
\begin{figure}[h]
  \centering
  \includegraphics[width=0.7\linewidth]{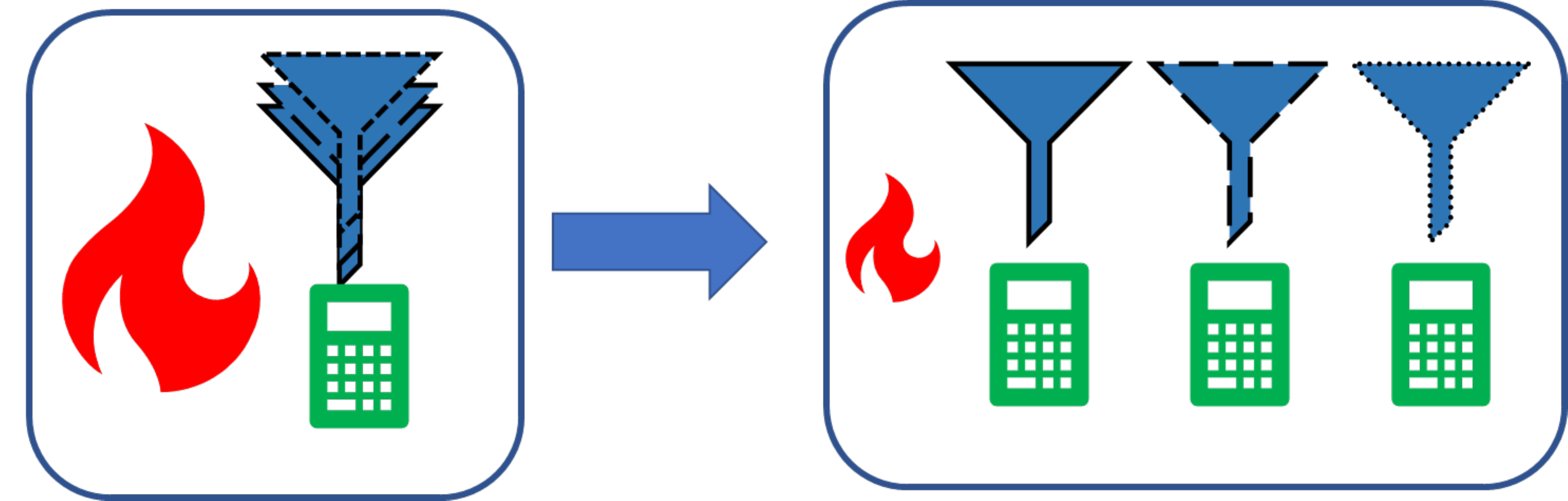}
  \caption{Traditional Scaling Bottlenecks and \proximus-style Scaling}
  \label{fig:summary_funnel_figure}
\end{figure}

As the amount of data created and analyzed grows at an exponential pace, OEMs
are willing to harness all forms of available compute to tackle
their computational needs. While CPUs are the dominant platform-of-choice for
DNN inference in datacenters, our in-depth analysis reveals significant
opportunities for more efficient CPU resource utilization leading to drastic
improvements in performance/watt and the ability to scale performance without
increasing cache capacity or bandwidth. Using a combination of simple ISA
enhancements and light-weight tensor compute \textit{near all caches} in the
CPU, \proximusspace raises the bar for CPU-based DNN-inference performance and
performance/watt while maintaining the same programming and memory models.
\proximusspace represents a fundamental re-imagination of the general-purpose
CPU, better suited for the AI era.

\bibliographystyle{IEEEtranS}
\bibliography{references}

\end{document}